\documentclass[12pt,a4paper,sort&compress]{article}

\pdfoutput=1

\usepackage{jheppub}
\usepackage{amsmath}
\usepackage{amssymb}
\usepackage{amsfonts}
\usepackage{amsthm}
\usepackage{mathtools}
\usepackage{stmaryrd}
\usepackage{shuffle}
\usepackage{enumitem}
\usepackage{color}   
\usepackage{hyperref}
\usepackage{adjustbox}
\usepackage{graphicx}
\usepackage[dvipsnames]{xcolor}
\colorlet{LightAquamarin}{Aquamarine!30}
\colorlet{LightRubineRed}{RubineRed!30}
\definecolor{UniBlau}{RGB}{7,82,154}
\usepackage{framed}
\usepackage{tikz}
\usepackage{bbold}

\newcommand{\rd}{\mathrm{d}}

\def\beq{\begin{equation}}
\def\eeq{\end{equation}}
\def\bsp#1\esp{\begin{split}#1\end{split}}

\newenvironment{sloppyequation}[0]{\sloppy\begin{flushleft}\hspace*{0.75cm}\(}{\)\end{flushleft}\fussy}

\newcommand{\beqsloppy}{\begin{sloppyequation}}
\newcommand{\eeqsloppy}{\end{sloppyequation}}

\newcommand{\cF}{\begin{cal}F\end{cal}}

\newcommand{\cL}{\begin{cal}L\end{cal}}

\newcommand{\bfY}{{\bf Y}}

\usepackage{hyperref}
\theoremstyle{definition}

\newtheorem{thm}{Theorem}

\newcommand{\bP}{{\textbf P}}
\newcommand{\bH}{{\textbf H}}
\newcommand{\bC}{{\textbf C}}
\newcommand{\bA}{{\textbf A}}
\newcommand{\bM}{{\textbf M}}

\newcommand{\tI}{{\text{I}}}

\newcommand{\nablasub}[1]{\nabla_{\!#1}}
\DeclareMathOperator{\Gr}{Gr}

\allowdisplaybreaks

\title{Feynman integrals in two dimensions and single-valued hypergeometric functions}

\author[a]{Claude Duhr}
\author[a]{Franziska Porkert}

\affiliation[a]{Bethe Center for Theoretical Physics, Universit\"at Bonn, D-53115, Germany}

\emailAdd{cduhr@uni-bonn.de}
\emailAdd{fporker@uni-bonn.de}

\abstract{We show that all Feynman integrals in two Euclidean dimensions with massless propagators and arbitrary non-integer propagator powers can be expressed in terms of single-valued analogues of Aomoto-Gelfand hypergeometric functions. The latter can themselves be written as bilinears of hypergeometric functions, with coefficients that are intersection numbers in a twisted homology group. As an application, we show that all one-loop integrals in two dimensions with massless propagators can be written in terms of Lauricella $F_D^{(r)}$ functions, while the $L$-loop ladder integrals are related to the generalised hypergeometric ${}_{L+1}F_L$ functions.}

\begin{document}

\preprint{BONN-TH-2023-09}

\maketitle

\flushbottom


\section{Introduction}
Feynman integrals occupy a central role in quantum field theory as the building blocks for perturbative calculations of scattering amplitudes. In particular, they connect theoretical predictions to the observables measured in today's large scale experiments -- particle colliders as well as gravitational wave detectors. As a result, substantial progress has been made in finding more efficient and precise methods for computing Feynman integrals, including a better understanding of their analytic structure. The study of these structures has lead to exciting connections with different areas of mathematics.
The simplest cases, which involve multiple polylogarithms and their extensions to elliptic curves, are by know relatively well understood (see, e.g., refs.~\cite{Weinzierl:2022eaz,Abreu:2022mfk,Bourjaily:2022bwx} and references therein for recent reviews). More recently, it was realised that also functions associated to Calabi-Yau varieties arise from Feynman integrals~\cite{Brown:2010bw,Bloch:2014qca,MR3780269,Bourjaily:2018yfy,Bourjaily:2019hmc, Klemm:2019dbm,vergu_traintrack_2020,Bonisch:2020qmm, Bonisch:2021yfw, Duhr:2022pch,Duhr:2022dxb,Pogel:2022yat, Pogel:2022ken,McLeod:2023doa,gorges_procedure_2023}.  Higher genus surfaces, formerly only considered for string amplitudes, are nowadays also studied in relation to Feynman integrals \cite{Huang:2013kh,Hauenstein:2014mda,marzucca2023genus}. All these functions typically arise in the $\epsilon$-expansion in dimensional regularisation, and it is known that the coefficients of this expansion are periods in the sense of Kontsevich and Zagier~\cite{MR1852188,Bogner:2007mn}. If the expansion in $\epsilon$ is not performed, the appropriate mathematical setting is twisted cohomology theory introduced by Aomoto (cf., e.g., ref.~\cite{aomoto_theory_2011}). In particular, it was shown that the intersection pairing on the twisted de Rham cohomology groups may be used as an alternative approach to perform the reduction to a basis of master integrals, cf., e.g., 
refs.~\cite{Mizera:2017rqa,Mastrolia:2018uzb,Mizera:2019ose,Frellesvig:2019kgj,Mizera_2020,Frellesvig:2019uqt,Mizera:2019vvs,Frellesvig:2020qot,Weinzierl:2020gda,Caron-Huot:2021xqj,Caron-Huot:2021iev,Chestnov:2022xsy,Cacciatori:2022mbi,Giroux:2022wav,Giroux:2023syp,Fontana:2023amt,De:2023xue}. 
Despite all this progress, a general understanding of the mathematical framework and the class of functions relevant for Feynman integrals is still lacking.

In this paper we show that twisted cohomology can also be used to compute some classes of Feynman integrals in two integer dimensions, without the need to appeal to dimensional regularisation. More specifically, we will consider Feynman integrals in two Euclidean dimensions with massless propagators and non-integer propagator exponents. It is well known that in two dimensions it is convenient to parametrise the two-dimensional external momenta or points in terms of complex variables (and their complex conjugates). In these variables the integrand becomes holomorphically separable, i.e., it is a product of a holomorphic and an anti-holomorphic function. We will show that this factorisation has a natural interpretation in the framework of twisted cohomology. Inspired by the work of Brown and Dupont on how to interpret closed string amplitudes at tree-level as the single-valued analogues of the corresponding open string amplitudes~\cite{Brown:2019wna,Brown:2018omk,Britto:2021prf}, we find that all Feynman integrals in two Euclidean dimensions with massless propagators and non-integer propagator exponents evaluate to single-valued analogues of hypergeometric functions. Using the double-copy formula derived in the context of string amplitudes, we can write these single-valued analogues of hypergeometric functions as bilinears of ordinary hypergeometric functions. The latter are well-studied in the mathematics literature, see, e.g., refs.~\cite{MR0841131,aomotoold,aomoto_theory_2011,yoshida_hypergeometric_1997,matsubaraheo2023lectures,kita_intersection_1994-2,kita_intersection_2006-2,yoshida_intersection_2000-1,kita_intersection_2006-4,mimachi_intersection_2010,goto2020lauricellas,GOTO_2013,goto2014intersection}. While this is not the first time that it is observed that Feynman integrals in two dimensions can be written as single-valued bilinears in hypergeometric functions~\cite{Mimachi_2003,derkachov_basso-dixon_2019,Duhr:2022pch,Derkachev:2022lay,Halder:2023nlp}, our results show that this is a completely general phenomenon that applies to all Feynman integrals in two dimensions. In particular, it is not restricted to position space integrals in two-dimensional conformal field theories (CFTs).


The paper is organised as follows: In section \ref{introfeyn} we introduce the class of Feynman integrals we are interested in. In section \ref{sec.twist}, we give a review of twisted cohomology, and in section \ref{sec:hypgeo} we explain how the so-called Aomoto-Gelfand hypergeometric functions can be treated in this framework and how one can construct their single-valued versions. Finally, in section \ref{SVPF} we give our main result, and we explain how every Feynman integral in two dimensions with massless propagators can be given as such a single-valued version of a hypergeometric function. We illustrate this on several examples, in particular all one-loop integrals and one-parameter ladder integrals with generic propagator exponents. We include several appendices where we review some of the mathematical background relevant to evaluate homology intersection numbers, which are needed to compute the single-valued hypergeometric functions. We also include an appendix where we show how to extend our findings in some weaker form to Feynman integrals in $D=1$ dimensions.
\section{Feynman integrals in two dimensions} 
\label{introfeyn}
 
In this section we introduce our conventions. Our main objects of interest are scalar Feynman integrals with massless propagators. In $D$ Euclidean space-time dimensions they can be defined as follows:
 \begin{align}
  \label{eq:Feynman_int_def}
 &\tI_G^{D}(\underline{\nu},\underline{v}) =\int\left(\prod_{j=1}^L \frac{\rd^D u_j}{\pi^{D/2}}\right) \prod_{j=1}^E\frac{1}{D_j(\underline{u},\underline{v})^{\nu_j}}\, .
  \end{align}
Here $G$ denotes a Feynman graph with $L$ loops and $E$ internal edges and $\underline{\nu}=(\nu_1,\ldots,\nu_{E})\in \mathbb{C}^E$ are the exponents of the propagators. This definition applies to  Feynman integrals in both momentum and position space, and $\underline{v}$ are the external momenta (if we work in momentum space) or external points (if we work in position space), respectively. It also includes numerators, which correspond to propagators raised to negative integer powers. We work in Euclidean kinematics, and the inverse propagators take the form
\beq\label{eq:props_form}
D_j(\underline{u},\underline{v}) = \Bigg(\sum_{i=1}^L\alpha_{ji}u_i+\sum_{i}\beta_{ji}v_i\Bigg)^2\,,\qquad \alpha_{ji},\beta_{ji}\in\{-1,0,1\}\,.
\eeq
Feynman integrals typically diverge and require regularisation. The most commonly used regularisation is dimensional regularisation, where the integrals are computed for non-integer space-time dimension $D$. Alternatively, one may consider the integrals as functions of the propagator exponents $\underline{\nu}$. Typically the integral converges for some generic values of $\underline{\nu}$ even for integer values of the dimension $D$, cf.,~e.g.,~refs.~\cite{71d79a7d-e98f-3da8-8de1-e3b1ba5446fc,Speer1975}.
In the remainder of this paper, we will be concerned with Feynman integrals with massless propagators in $D=2$ dimensions and with generic propagator powers, so that we will not need to discuss the convergence of the integrals. The case of integer powers can be recovered by expanding around the appropriate integer values of $\nu_i$. 

For $D= 2$, the  expression for the Feynman integral in eq.~\eqref{eq:Feynman_int_def} can be cast in a different form. 
The integration variables $u_i$ as well as the external data $v_i$ are all two-dimensional real vectors. It is convenient to package the two real components of such a vector into a single complex variable:
  \begin{align}
 x_j =u_j^1+iu_j^2 \text{ and } y_j = v_j^1+ iv_j^2\,.
 \end{align}
 Since we work in Euclidean kinematics, the propagators can be written as
\beq\label{eq:props_2D}
D_j(\underline{u},\underline{v}) = \big|L_j(\underline{x},\underline{y})\big|^2\,,\qquad L_j(\underline{x},\underline{y}) = \sum_{i=1}^L\alpha_{ji}x_i+\sum_{i}\beta_{ji}y_i\,.
\eeq
It is easy to check that the Feynman integral then takes the form
 \begin{align}
 \label{EQU.762}
 &\tI_G(\underline{\nu},\underline{y}) = \left(-\frac{1}{2\pi i}\right)^L\int_{\mathbb{C}^L} \left(\bigwedge_{j=1}^L  \rd x_j\wedge \rd \bar{x}_j \right)\prod_{j=1}^E\frac{1}{|L_{j}(\underline{x},\underline{y})|^{2\nu_j}} \, ,
  \end{align}
  where we drop the dependence of the integral on the dimension $D=2$. Let us make some comments about this way of writing a Feynman integral in two dimensions. First, we stress that it is essential that the propagators are massless, because otherwise the inverse propagators cannot be written as absolute values squared. Second, since the $L_{j}(\underline{x},\underline{y})$ are linear, we see that in the holomorphic variables $(\underline{x},\underline{y})$, the propagator singularities are supported on the hyperplanes $L_{j}(\underline{x},\underline{y})=0$ in complex affine space $\mathbb{C}^L$. Finally, we see that the integrand is holomorphically separable, i.e., it can be written as a product of a holomorphic and an anti-holomorphic function. We can make this explicit by introducing a holomorphic $L$-form $\omega_G$ defined by
  \beq
  \label{omegag}
  \omega_G = \left(\bigwedge_{j=1}^L  \rd x_j \right)\prod_{j=1}^E\frac{1}{L_{j}(\underline{x},\underline{y})^{\nu_j}}\,.
  \eeq
  The Feynman integral can then be written as
  \beq
  \label{eq:2DIntegral}
 \tI_G(\underline{\nu},\underline{y}) = (-1)^{\frac{L(L-1)}{2}} \left(-\frac{1}{2\pi i}\right)^L
 \int_{\mathbb{C}^L} \omega_G\wedge \overline{\omega}_G
 \,.
 \eeq
Note that $\omega_G$ implicitly depends on $\underline{\nu}$. Unless all the entries of $\underline{\nu}$ are integers, $\omega_G$ is not a rational differential form, but it is multi-valued. Owing to the absolute value in eq.~\eqref{eq:props_2D}, the propagators and  the $(L,L)$-form $\omega_G\wedge\overline{\omega}_G$ are single-valued. In order to understand the class of integrals that arise from Feynman integrals in two dimensions, we therefore need to understand the mathematics underlying integrals arising from such differential forms. The correct mathematical setting for integrands containing multi-valued forms is twisted cohomology, which we will review in the next section.

\section{Review of twisted cohomology}
\label{sec.twist}

In this section we review the mathematical framework used throughout the rest of this paper. For a more detailed overview of twisted cohomology, specifically applied to hypergeometric functions, see,~e.g.,~refs.~\cite{aomoto_theory_2011,yoshida_hypergeometric_1997,matsubaraheo2023lectures}.
  
\subsection{General setup} 

We are interested in integrals of the form 
\begin{align}
\label{EQU.5}
 \int_\gamma \Phi \, \varphi \, , 
\end{align}
where $\varphi := \varphi(\underline{x})\,\rd x_1\wedge \dots \wedge \rd x_n$ is a rational differential $n$-form on the space $X=\mathbb{C}^n-\Sigma$ parametrised by affine coordinates $(x_1,\dots, x_n)$. $\gamma$ is an integration contour on $X$ with vanishing boundary, $\partial\gamma=0$ (we will be more precise below). The function
\begin{align}
\label{EQU.6}
\Phi= \prod_{i=0}^r L_i(\underline{x})^{a_i}\, ,\qquad L_i(\underline{x}) = \rho_{i0}+\sum_{j=1}^n\rho_{ij}x_j\,,\qquad \rho_{ij}\in\mathbb{C}\,,
\end{align}
is the so-called \textit{twist}. Throughout this paper we will assume that $L_i(\underline{x})$ is a linear form, and so each $L_i(\underline{x})=0$ defines a hyperplane in $\mathbb{C}$.
$\Sigma$ is the variety defined by the union of these hyperplanes:
\begin{align}
\label{EQU.10}
\Sigma =\bigcup_{i} \{L_i(\underline{x})=0\}\, . 
\end{align}
We generally assume that at least one of the exponents $a_i$ is non-integer, making $\Phi$ a multi-valued function.  Specifically, in this paper we take all the $a_i$ to be non-integer, real and negative. Other values can be obtained by taking appropriate limits or performing appropriate analytic continuations. We assume that the twist is defined such that the singularities of the rational forms $\varphi$ also lie in $\Sigma$. The integral may have a singularity at infinity, and we associate to the hyperplane at infinity the exponent $a_{\infty}=-\sum_ia_i$. Let us define a family of integrals by fixing a twist $\Phi$. We want to work modulo forms that vanish upon integration. By Stoke's theorem these are exact forms $\rd \!\left(\Phi \tilde{\varphi}\right)$, as 
\begin{align}
    \int_\Gamma \rd \!\left(\Phi \tilde{\varphi}\right) = \int_{\partial \Gamma} \Phi \tilde{\varphi} =0 \,.
\end{align}
where the last step follows from $\partial\Gamma=0$. So we want to introduce the equivalence relation $\Phi \varphi \sim \Phi \varphi+ \rd (\Phi \tilde{\varphi})$. Since we  consider the twist to be fixed, it is more convenient to define the equivalence relation on the level of the rational forms $\varphi$. If $\tilde{\varphi}$ is a rational differential form, we have:
\begin{align}
\label{EQU.7}
\rd \!\left(\Phi \tilde{\varphi}\right)=  \Phi \,\rd \tilde{\varphi}+\rd \Phi \wedge \tilde{\varphi} = \Phi \left(\rd \tilde{\varphi}+\frac{\rd \Phi}{\Phi} \wedge \tilde{\varphi} \right)= \Phi\, \nablasub{\underline{a}} \tilde{\varphi}\,,
\end{align}
where we defined 
\begin{align}
\label{EQU.8}
\nablasub{\underline{a}} = \rd + \Omega \wedge \cdot \text{~~~with~~~} \Omega = \frac{\rd\Phi}{\Phi} = \rd\!\log \Phi \, . 
\end{align}
Thus, we want to consider the equivalence relation $\varphi\sim \varphi + \nablasub{\underline{a}} \tilde{\varphi}$. We note that $\nablasub{\underline{a}}^2=0$, i.e., $\nablasub{\underline{a}}$ defines a flat connection. The dual connection is $\check{\nabla}_{\!\underline{a}}= \nablasub{-\underline{a}}$. The natural framework for working with closed forms modulo exact forms in the presence of a multi-valued twist is to work with representatives of twisted cohomology groups: 
\begin{align}
\label{EQU.11}
H_{\textrm{dR}}^k \left(X, \nablasub{\underline{a}} \right) =\{ \text{$k$-forms $\varphi$ on $X$ with $\nablasub{\underline{a}} \varphi =0$}\}/ \{ \text{exact forms } \nablasub{\underline{a}} \tilde{\varphi} \text{ on $X$}\}\, . 
\end{align}
The $k$-forms $\varphi$ on $X$ with $\nablasub{\underline{a}} \varphi =0$ are the so-called \textit{twisted co-cycles}.  $H_{\textrm{dR}}^k\left(X, \nablasub{\underline{a}} \right) $ is always finite-dimensional and $H^k_{\text{dR}}\left(X, \nablasub{\underline{a}} \right)=0$ for $k\neq n$, cf.,~e.g.,~ref.~\cite{aomoto_theory_2011}.
We will therefore only focus on the case $k=n$ from now on. 
One can also define a twisted de Rham cohomology group with twist $\Phi^{-1}$, and the corresponding dual connection $\check\nabla_{\!\underline{a}}$:
\begin{align}
\label{EQU.13.A}
H_{\textrm{dR}}^n(X, \check{\nabla}_{\!\underline{a}} ) &=\{ \text{$n$-forms $\check{\varphi}$ on $X$ with $\check{\nabla}_{\!\underline{a}} \check{\varphi} =0$}\}/ \{ \text{exact forms }\check{\nabla}_{\!\underline{a}}\check{\tilde{\varphi}}^{}\text{ on $X$}\} \, .
\end{align}
In general $X$ is not compact and the forms $\check{\varphi}\in H_{\text{dR}}^n(X, \check{\nabla}_{\!\underline{a}})$ are not necessarily compactly supported. Though, $H_{\text{dR}}^n(X, \check{\nabla}_{\!\underline{a}})$ is isomorphic to a compactly supported de Rham cohomology group:
\begin{align*}
H_{\textrm{dR},c}^n(X, \check{\nabla}_{\!\underline{a}} ) &=\{ \text{compactly supported $n$-forms $\check{\varphi^{}}_c$ with $\check{\nabla}_{\!\underline{a}} \check{\varphi}_c =0$}\}/ \{ \text{exact forms} 
\} \, .
\end{align*}
An explicit construction of the isomorphism $\iota_c: H^n_{\text{dR}}\rightarrow H^n_{\text{dR},c},\check{\varphi}\mapsto\check{\varphi}_c$ can be found in refs.~\cite{Caron-Huot:2021iev,Caron-Huot:2021xqj}. We call the elements of $H_{\textrm{dR},c}^n(X, \check{\nabla}_{\!\underline{a}} )$ the \textit{dual twisted cocyles} and the duality is given by the intersection pairing in eq.~(\ref{EQU.19}).

For the integral in eq.~(\ref{EQU.7}) to be well-defined over any  form $\varphi\in H_{\textrm{dR}}^n \left(X, \nablasub{\underline{a}} \right)$, we require the  contour to be compact and we need to choose a branch of the twist on this contour. To specify this branch choice, we use the notation 
\begin{align}
\label{twistedcycel}
    \gamma= \eta_c \otimes \Phi|_{\eta_c}= \sum_\triangle^{\text{finite}} c_{\triangle} \triangle \otimes \Phi|_{\triangle} \, , 
\end{align}
where $\Phi|_{\eta_c}$ denotes the branch choice on $\eta_c$ from $\check{\mathcal{L}}_{\underline{a}}$,  the locally constant sheaf of $\check{\nabla}_{\underline{a}}$-horizontal sections generated by $\Phi$. The sum of embedded simplices $\triangle$ with coefficients $c_{\triangle}\in\mathbb{C}$ is finite, making $\eta_{c}$ compact.

We want to consider boundaryless contours and additionally work modulo boundaries (similar to how we work with closed modulo exact forms from twisted cohomology groups). To make this prescription more precise we consider the contours as representatives of a so-called twisted homology group 
\begin{align}
     H_{n}(X, {\check{\mathcal{L}}}_{\underline{a}}) = \{ \gamma=\eta_c\otimes \Phi|_{\eta_c} \text{ with } \partial_{\underline{a}}\gamma =0\}/\{ \text{ boundaries }\partial_{\underline{a}} \tilde{\gamma}\}\, , 
\end{align}
where $\partial_{\underline{a}}$ is the boundary operator that maps a contour $\eta_c$ to its boundary and restricts the twist loaded onto that contour to this boundary. The homology group $  H_{n}(X, {\check{\mathcal{L}}}_{\underline{a}})$ is Poincar\'e dual to a compactly supported twisted de Rham cohomology group: 
\begin{align}
    H_{\textrm{dR},c}^{n} (X, \check\nabla_{\!\underline{a}}) \cong H_{n}(X, {\check{\mathcal{L}}}_{\underline{a}}) \, . 
\end{align}
On the other hand, the twisted de-Rham cohomology group of eq.~(\ref{EQU.11}) is Poincar\'e-dual to a locally-finite twisted homology group \cite{aomoto_theory_2011}: 
\begin{align}
    \label{Poincare}
    H^n_{\textrm{dR}}(X, \nablasub{\underline{a}}) \cong H_{n}^{\text{lf}} (X, {\mathcal{L}}_{\underline{a}})\, . 
\end{align}
The elements of $H_{n}^{\text{lf}} (X, {\mathcal{L}}_{\underline{a}})$ take the form 
\begin{align}
\label{duatwistedcycle}
    \check{\gamma} =\eta \otimes \Phi^{-1}|_{\eta}=\sum_{\square}^{\text{locally finite}} d_\square \square \otimes \Phi^{-1}|_{\square}\, . 
\end{align}
 The linear combinations of eq.~(\ref{duatwistedcycle})  are only required to be \emph{locally finite}, i.e., they may be infinite, but every compact set of $X$ only intersects a finite number of summands. Locally finite twisted cycles can  naturally be paired with a differential form $\check{\varphi}_c$ with compact support, because only a finite number of summands intersect the support of $\check{\varphi}_c$. The branch $\Phi^{-1}|_{\eta}$  is taken from $\mathcal{L}_{\underline{a}}$, the sheaf of $\nablasub{\underline{a}}$-local sections generated by $\Phi^{-1}$ and $\check{\mathcal{L}}_{\underline{a}} \cong \mathcal{L}_{-\underline{a}}$. The regularisation procedure to map a contour $\eta$ to its compactly supported version $\eta_c$ is described in appendix \ref{sec.hom.1}  for the case of a twist with linear factors along with several examples focusing on the univariate case. 

All these (co-)homology group are finite-dimensional. It is known how to find bases for the corresponding spaces, cf.,~e.g.,~refs.~\cite{aomoto_theory_2011,Mizera:2017rqa,Mizera:2019ose}. In particular, in the case where all the hyperplanes are in general position and all exponents $a_i$ are non-integer (including $a_{\infty}$), a basis for $H_{\text{dR}}^n (X, \nablasub{\underline{a}}) $ is obtained by considering $n$-forms with logarithmic singularities on $L_i(\underline{x})=0$, while a basis for $H_n^{\text{lf}}(X,\mathcal{L}_{\underline{a}}) $ is given by considering the bounded chambers with boundaries in $\Sigma$.

\subsection{Pairings between (co-)homology groups}

In order to discuss bilinear pairings between the (co-)homology groups, it is convenient to introduce a 
bra-ket-notation for twisted (co-)cycles and their duals. We use the following notations:
\begin{itemize}
\item $\langle \varphi |$ denotes a twisted co-cycle in $H_{\text{dR}}^n (X, \nablasub{\underline{a}}) $,
\item $|\check{\varphi}\rangle $ denotes a dual twisted  co-cycle  in $H_{\text{dR},c}^n (X, \check{\nabla}_{\!\underline{a}}) $, 
\item $|\gamma ]$ denotes a twisted cycle in $H_n (X,\check{\mathcal{L}}_{\underline{a}}) $, 
\item $[\check{\gamma}|$ denotes a (locally-finite) dual twisted cycle  in $H_n^{\text{lf}} (X,{\mathcal{L}}_{\underline{a}}) $. 
\end{itemize}
In the following, it will be useful to fix bases $\{\langle\varphi_i |\}$, $\{| \check{\varphi}_i^{}\rangle\}$, $\{|\gamma_i]\}$ and $\{[\check{\gamma}_i^{}|\}$ of $H_{\textrm{dR}}^n(X, \nablasub{\underline{a}})$, $H_{\textrm{dR},c}^n(X, \check{\nabla}_{\!\underline{a}})$, $H_n(X, \check{\mathcal{L}}_{\underline{a}})$ and $H_n^{\text{lf}} (X,{\mathcal{L}}_{\underline{a}})$ respectively. We also implicitly assume that the basis and the dual basis are related to each other by  regularisation procedures and not chosen independently in both cases. Using this notation, we can define four different non-degenerate pairings:

\begin{enumerate}
\item \underline{The (twisted) period pairing $\langle .|.]$} between a twisted cycle $|\gamma]$ and a twisted cocylce $\langle  \varphi |$ is defined by
\begin{align}
\label{EQU.15}
\langle  \varphi |\gamma] = \int_{\gamma} \Phi\varphi\, . 
\end{align}
An integral as in eq.~\eqref{EQU.15} is called a \emph{twisted period}.
Note that the convergence of this integral may put conditions on the allowed values of $\underline{a}$. If we see the integral as a function of $\underline{a}$, the integral is typically convergent in some region, and other values can be reached by appropriate limits or analytic continuation. We will therefore not discuss the convergence of twisted period integrals in the following.

It will be useful to consider the matrix $\bP_{\!\!\underline{a}}$ -- called the \emph{(twisted) period matrix} -- formed by the twisted periods obtained by pairing the basis elements (for a fixed choice of basis):
\begin{align}
\label{EQU.16}
(\bP_{\!\!\underline{a}})_{ij} = \langle \varphi_i |\gamma_j] = \int_{\gamma_j} \Phi\varphi_i\, .
\end{align}
\item \underline{The {dual period} pairing $[.|.\rangle$} between a twisted dual cycle $[\check{\gamma}^{}|$ and a twisted dual cocyle $|\check{\varphi}^{}\rangle$ is defined by
\begin{align}
\label{EQU.18}
[\check{\gamma}^{}|  \check{\varphi}^{}\rangle = \int_{\check{\gamma^{}}} \Phi^{-1}\iota_c(\check{\varphi})\, . 
\end{align}
The map $\iota_c$ replaces $\check{\varphi}$ with a representative of the cohomology class $[{\check{\varphi}}]$ from $ H_{\textrm{dR},c}^{n} (X, \check{\nabla}_{\underline{a}}) $ that has compact support. Details on this map are given in \cite{Caron-Huot:2021iev,Caron-Huot:2021xqj,Giroux:2022wav}. Considering the pairing for a fixed choice of basis defines the dual twisted period matrix $\check{\bP}_{\!\!\underline{a}}$ with entries
$(\check{\bP}_{\!\!\underline{a}})_{ij}=[\check{\gamma}_j^{}|\check{\varphi}_i^{}\rangle$. Since the only difference between eqs.~\eqref{EQU.16} and~\eqref{EQU.18} lies in the twist, one can show that we have the relation $\check{\bP}_{\!\!\underline{a}} = \bP_{\!\!-\underline{a}}^T$.
\item \underline{The {intersection} pairing $\langle .|.\rangle$} between twisted cocylces $\varphi_A, \check{\varphi}_B$ is defined by
\begin{align}
\label{EQU.19}
\langle \varphi_A| \check{\varphi}_B^{}\rangle =\int_{X}\varphi\wedge \iota_c(\check{\varphi}_B) \, . 
\end{align}
 Methods for the computation of the intersection matrix of cocycles have for example been developed in refs.~\cite{matsumoto_intersection_k-forms,aomoto_theory_2011,Mizera:2017rqa,Mizera:2019ose,Frellesvig:2019uqt,Frellesvig:2020qot,Mizera:2019vvs}. 
We define the \textit{intersection matrix} for our choice of bases:
\begin{align}
\label{EQU.20}
(\bC_{\underline{a}})_{ij} = \frac{1}{(2\pi i)^n}\langle \varphi_i |\check{\varphi}_j^{} \rangle =  \frac{1}{(2\pi i)^n}\int_X \varphi_i \wedge \iota_c(\check{\varphi}_j^{})\,.
\end{align}
\item \underline{The {intersection} pairing $[.|.]$} between a twisted cycle and a dual twisted cycle can be thought of as counting the intersections between these cycles taking into account the branch choice of the twist near these intersections. For more details we refer to the literature~\cite{aomoto_theory_2011,yoshida_hypergeometric_1997}.  Since the intersection matrix of the homology group is central to this paper, we also give a  review of its computation in appendix \ref{sec.hom.2}.  In all cases of interest for this paper, the intersection numbers evaluate to rational functions (with rational coefficients) of $e^{2\pi ia_j}$. We define an intersection matrix for the bases of the homology group and its dual: 
 \begin{align}
     (\bH_{\underline{a}})_{ij} = [\check{\gamma}_j|\gamma_i]\, . 
 \end{align}
\end{enumerate} 

These pairings are perfect, i.e. they allow us to establish completeness relations for the  homology and the cohomology groups respectively
\beq\bsp
    \label{complet.1}
    \mathbb{1} &\,= \left(2\pi i\right)^{-n} |\check{\varphi}_i \rangle \left(\bC_{\underline{a}}^{-1}\right)_{ij} \langle \varphi_j |\,, \\
    \mathbb{1} &\,=  |\gamma_i ] \left(\bH_{\underline{a}}^{-1}\right)_{ji} [ \check{\gamma}_j | \, .
\esp\eeq
Inserting a completeness relation into the homology and cohomology intersection pairings we obtain the \textit{twisted Riemann bilinear relations}~\cite{cho_intersection_1995}: 
\beq\bsp
\label{EQU.23}
[\check{\gamma}_A|\gamma_B]&\,=(2\pi i)^{-n} \,[\check{\gamma}_A |\varphi_i^{} \rangle\left(\bC_{\underline{a}}^{-1}\right)_{ij} \langle \varphi_{j} |\gamma_B]\,, \\
\langle\varphi_A|\check{\varphi}_B\rangle&\,= \langle\varphi_A|\gamma_i^{} ]\left(\bH_{\underline{a}}^{-1}\right)_{ji} [ \gamma_{j}|\check{\varphi}_B\rangle\, .
\esp\eeq
We can cast the twisted Riemann bilinear relations into relations between matrices:
\beq\bsp
\label{EQU.24}
\bH_{\underline{a}} &\,= (2\pi i)^{-n}\, \bP_{\!\!\underline{a}}^T\, (\bC^{-1}_{\underline{a}})^{T}\,\bP_{\!\!-\underline{a}} \,, \\ 
\bC_{\underline{a}} &\,= (2\pi i)^{-n} \,\bP_{\!\!\underline{a}}\, (\bH^{-1}_{\underline{a}})^{T}\,\bP_{\!\!-\underline{a}}^T
\, . 
\esp\eeq
Note that from this it follows that the inverse of the twisted period matrix is related to the transpose of the dual twisted period matrix:
\begin{align}
\label{EQU.25}
\bP_{\!\!\underline{a}}^{-1} = (2\pi i)^{-n}\,(\bH^{-1}_{\underline{a}})^{T}\,\bP_{\!\!-\underline{a}}^T\, \bC_{\underline{a}}^{-1} \, . 
\end{align}

\subsection{Example: Euler's Beta function}
\label{sec:beta_example}
We now illustrate the concepts from the previous section on a simple example, namely \textit{Eulers's Beta  function}:
\begin{align}
\label{beta1}
\beta(\alpha_1,\alpha_2)=  \int_0^{1} x^{\alpha_1-1} \left(1-x \right)^{\alpha_2-1} \rd x=\frac{\Gamma(\alpha_1)\Gamma(\alpha_2)}{\Gamma(\alpha_1+\alpha_2)}\, .  
\end{align}
Euler's beta function has already been discussed several times in the literature in the context of twisted cohomology theories. We nevertheless reproduce this example here, as it is the simplest case and will have an interpretation in the context of Feynman integrals in two dimensions later on.

Let us define $X:=\mathbb{C}-\{0,1\}$. From the integrand in eq.~\eqref{beta1}, we see that the twist is
\begin{align}
\label{twistbeta}
\Phi=\frac{1}{x^{a_1} (1-x)^{a_2}}
\end{align} 
with $a_1,a_2$ the non-integer parts of $\alpha_1,\alpha_2$ and $\underline{a}=(a_1,a_2)$.
The cohomology group $H_{\textrm{dR}}^1(X, \nablasub{\underline{a}}) $ is one-dimensional. As a basis element we choose the cocyle $\varphi=\frac{\rd x}{x(1-x)}$. The intersection matrix is
\beq
\label{intbeta1}
\bC_{\underline{a}}= \begin{pmatrix}\langle\varphi_{}|\check{\varphi}_{}\rangle\end{pmatrix}= \begin{pmatrix}\frac{a_1+a_2}{a_1a_2}\end{pmatrix}\,.
\eeq
The corresponding homology group $H_1(X_, \check{\mathcal{L}}_{\underline{a}})$ is spanned by $\gamma=\eta_{c} \otimes \Phi|_{\eta_{c}}$, where $\eta_{c}$ denotes the regularised version of the open interval $(0,1)$ with the regularisation procedure described in appendix \ref{sec.hom.1}. The intersection matrix is:
\beq\bsp
 \label{EQU.83}
\bH_{\underline{a}} &\,=\begin{pmatrix}[\check{\gamma}_{}|\gamma_{}]\end{pmatrix} = \begin{pmatrix}-\frac{e^{2\pi ia_1}e^{2\pi ia_2}-1}{\left(e^{2\pi i a_1}-1\right)\left(e^{2\pi i a_2}-1\right)}\end{pmatrix}\\
&\,= \begin{pmatrix}-\frac{1}{2\pi i} \frac{\Gamma(1-a_1) \Gamma(a_1)\Gamma(1-a_2)\Gamma(a_2)}{\Gamma(1-a_{12}) \Gamma(a_{12})}\end{pmatrix}  \, ,
\esp\eeq
with $a_{12}=a_1+a_2$. Finally, the period matrix is 
 \begin{align}
 \label{EQU.82}
\bP_{\underline{a}}=\begin{pmatrix}\langle \varphi|\gamma]\end{pmatrix}= \begin{pmatrix} \beta(a_1,a_2)\end{pmatrix}\, . 
 \end{align}
Note that this period matrix together with the intersection matrices in eqs.~(\ref{EQU.83}) and (\ref{intbeta1}) fulfil the twisted Riemann bilinear relations of eq.~(\ref{EQU.24}).

\section{Hypergeometric functions and their single-valued analogues} 
\label{sec:hypgeo}

\subsection{Aomoto-Gelfand hypergeometric functions}
\label{amgelhyp}

So far we have assumed that the coefficients $\rho_{ij}$ that define the linear forms $L_j(\underline{x})$ in eq.~(\ref{EQU.6}) are constant. We will now relax this assumption and look at families of twisted periods that are functions of these coefficients. This will lead to a very general class of hypergeometric functions first introduced by Gelfand~\cite{MR0841131,gelfand_algebraic_1986} and then further explored by Aomoto~\cite{aomotoold,aomoto_theory_2011}. In the remainder of this section we give a short review of these so-called \emph{Aomoto-Gelfand hypergeometric functions}.

An Aomoto-Gelfand hypergeometric function of type $(n+1,r+1)$ ($n$ and $r$ are integers) has a multi-valued integral representation of the form 
\begin{align}
\label{EQU.107} 
\cF(\underline{\alpha},{\bfY}) = \int_\Gamma
\Phi\,\omega= \int_\Gamma
\Phi\,f\,\omega_n\,.
\end{align} 
Here $\underline{\alpha}=(\alpha_0,\ldots,\alpha_r)$ is an $(r+1)$-dimensional vector and $\bfY$ is an $(n+1)\times (r+1)$ matrix:
\begin{align}
\label{EQU.106} 
{\bfY} = \begin{pmatrix} Y_{00} &\dots &Y_{0r} \\ 
\vdots &\ddots&\vdots \\ Y_{n0} & \dots&Y_{nr} \end{pmatrix} \, . 
\end{align}  
These data define the twist
\beq\bsp
\label{EQU.105} 
\Phi&\,= \prod_{j=0}^r \left(Y_{0j} \tau_0+Y_{1j} \tau_1+\dots+Y_{nj}\tau_n\right)^{a_j} \\
&\,= \prod_{j=0}^r \left(Y_{0j} +Y_{1j} x_1+\dots+Y_{nj}x_n\right)^{a_j}\, ,
\esp\eeq
where $a_i$ are the non-integer parts of $\alpha_i$, and $f$ is a rational function with poles at most at the location of the zeroes of the twist. In the following it will be useful to define $\underline{a} = (a_0,\ldots,a_r)$.
 $[\tau_0:\ldots:\tau_n]$ are homogeneous coordinates on $\mathbb{P}^n(\mathbb{C})$, and we introduced the affine coordinate chart $x_i=\frac{\tau_i}{\tau_0}$, $1\le i\le n$.
Moreover, the integrand involves the holomorphic $n$-form on $\mathbb{P}^n(\mathbb{C})$:
\beq\label{eq:topform}
\omega_n =  \sum_{i=0}^n (-1)^i \tau_i \,\rd \tau_0\wedge \dots \wedge\widehat{\rd \tau_{i}}\wedge\dots \wedge\rd \tau_n  = \bigwedge_{i=1}^n\rd x_i \,,
\eeq
where the hat indicates that the corresponding element has been omitted. If $\alpha_i=a_i$ for all $i$, $f=1$. The rational differential form $\omega=f\,\omega_n$ defines an element in the twisted cohomology group $H^n_{\text{dR}}(X_{\bfY},\nablasub{\underline{a}})$, where $X_{\bfY}$ is the $n$-dimensional complex space $\mathbb{C}^n$ with the hyperplanes $L_j(\underline{x}) = Y_{0j} +Y_{1j} x_1+\dots+Y_{nj}x_n=0$ removed. The integration region is supposed to be a fixed element $\Gamma\in H_n(X_{\bfY},{\check{\cL}}_{\underline{a}})$. We then see that an Aomoto-Gelfand hypergeometric function defines a family of twisted periods, $\cF(\underline{\alpha},{\bfY}) = \langle \omega|\Gamma]$. We can fix bases of the twisted cohomology and homology groups (we use the notations and conventions of the previous section) and decompose $\cF(\underline{\alpha},{\bfY})$ into that basis. We find
\beq
\cF(\underline{\alpha},{\bfY}) = \underline{c}^T\,{\bf C}_{\underline{a}}^{-1}\,{\bf P}_{\!\underline{a}}\,\big({\bf H}_{\underline{a}}^{-1}\big)^T\,\underline{h}\,,
\eeq
where we introduced the vectors 
\beq\bsp
\underline{c} &\,= (2\pi i)^{-n}\big(\langle\omega|\check{\varphi}_1\rangle,\ldots,\langle\omega|\check{\varphi}_d\rangle\big)^T\,,\\
\underline{h} &\,=\big([\check{\gamma}_1|\Gamma],\ldots,[\check{\gamma}_d|\Gamma]\big)^T\,,
\esp\eeq
where $d$ is the dimension of the (co-)homology group. We see that the information on $\cF(\underline{a},{\bfY})$ is encoded in the twisted period matrix ${\bf P}_{\!\underline{a}}$ (and the matrices of intersection numbers ${\bf C}_{\!\underline{a}}$ and ${\bf H}_{\!\underline{a}}$). The period matrix ${\bf P}_{\!\underline{a}}$ depends on ${\bfY}$, and it satisfies a first-order differential equation:
\beq\label{eq:GM}
\rd {\bf P}_{\!\underline{a}} = {\bf A}_{\underline{a}}\,{\bf P}_{\!\underline{a}}\,,
\eeq
where $\rd$ denotes the total differential and ${\bf A}_{\underline{a}}$ is a matrix of rational functions in $\bfY$. At this point we have to make a comment that will be important later on. The entries of $\bfY$ are complex numbers, and so the total differential has be to taken over all entries of $\bfY$ and its complex conjugate $\overline{\bfY}$. We can correspondingly split the total differential $\rd$ into its holomorphic and anti-holomorphic parts:
\beq
\rd = \partial + \overline{\partial} = \sum_{i,j}\rd Y_{ij}\,\frac{\partial}{\partial Y_{ij}}+\sum_{i,j}\rd \overline{Y}_{ij}\,\frac{\partial}{\partial \overline{Y}_{ij}}\,.
\eeq
The period matrix only depends on the holomorphic variables $Y_{ij}$, and so eq.~\eqref{eq:GM} can be cast in the equivalent form:
\beq\label{eq:GM_holom}
\partial {\bf P}_{\!\underline{a}} = {\bf A}_{\underline{a}}\,{\bf P}_{\!\underline{a}}\,.
\eeq

Note that $\bfY$ defines $r+1$ hyperplanes in $\mathbb{P}^n(\mathbb{C})$. One can show that $\cF(\underline{{\alpha}},\bfY)$ defines a hypergeometric function on the Grassmannian $\Gr(r+1,n+1)$. We will not discuss the general properties of Aomoto-Gelfand hypergeometric functions in detail here, but we refer to the literature (cf.,~e.g.,~refs.~\cite{MR0841131,gelfand_algebraic_1986,yoshida_hypergeometric_1997,aomotoold,aomoto_theory_2011}). They contain many well known classes of hypergeometric functions as special cases. In the following we discuss those examples that we will need for this work. As is commonly done in the literature, we will simply refer to the Aomoto-Gelfand hypergeometric functions as \emph{hypergeometric functions}.

\subsubsection{Lauricella functions $F_D^{(r)}$} 
\label{sec:Lauricella}

The \textit{Lauricella functions} $F_D^{(r)}$ are defined by the hypergeometric series
\begin{align}
F_D^{(r)}(\alpha, \underline\beta, \gamma; \underline{y})= \sum_{i_1,\dots, i_r=0}^\infty \frac{(\alpha)_{i_1+\dots+i_r} (\beta_1)_{i_1}\dots(\beta_r)_{i_r}}{(\gamma)_{i_1+\dots+ i_r} i_1!\dots i_r!} y_1^{i_1}\dots y_r^{i_r}\,,
\end{align}
where we defined $\underline\beta=(\beta_1,\dots, \beta_r)$ and $\underline{y}=(y_1,\dots, y_r)$. The Pochhammer symbol is defined by
\beq
(a)_n = \frac{\Gamma(a+n)}{\Gamma(a)}\,.
\eeq
The Lauricella $F_D^{(r)}$ admit the integral representation
\begin{align}
\label{EQU.108} 
F_D^{(r)}(\alpha, \underline{\beta},\gamma;\underline{y})= \frac{\Gamma(\gamma)}{\Gamma(\alpha)\Gamma(\gamma-\alpha)}\int_{0}^1x^{\alpha} (1-x)^{\gamma-\alpha}
 \prod_{j=1}^{r} (1-y_{j} x)^{-\beta_j}\frac{ \rd x}{x(1-x)}\, . 
\end{align} 
The singularities in $\Sigma_r=\{0,y_0^{-1}=1,y_1^{-1},\dots, y_{r}^{-1},\infty\}$ are distinct points in $\mathbb{C}$ and $({\alpha},\underline{\beta},\gamma)$ is a vector of complex numbers. Additionally, we define the normalised version of the Lauricella functions $F_D^{(r)}$ by
\begin{align}
\label{normLaur}
\mathcal{F}_D^{(r)}(\alpha, \underline{\beta},\gamma;\underline{y})=\beta(\gamma-\alpha, \alpha)\,F_D^{(r)}(\alpha, \underline{\beta},\gamma; \underline{y})\, .
\end{align}
From the previous equation, we can recognise them as Aomoto-Gelfand hypergeometric functions of type $(2, r+2)$ with 
\begin{align}
\label{EQU.109} 
{\bfY} = \begin{pmatrix} 0 & 1&1&\dots &1 \\
1&-1&- y_1&\dots &-y_{r} \end{pmatrix} \, .
\end{align}

For $r=0$, we recover the definition of the beta function discussed in section~\ref{sec:beta_example}, and for $r=1,2$ we obtain the well-known hypergeometric functions of Gauss and Appell:
\beq\bsp\label{eq:GaussAppell}
{}_2F_1(\alpha,\beta_1,\gamma;y_1) &\,= F_D^{(1)}(\alpha,\beta_1,\gamma;y_1)\,,\\
F_1(\alpha,\beta_1,\beta_2,\gamma;y_1,y_2) &\,= F_D^{(2)}(\alpha,\beta_1,\beta_2,\gamma;y_1,y_2)\,.
\esp\eeq
We also introduce the following notations for the normalised versions of Gauss' and Appell's hypergeometric functions:
\begin{align}\label{eq:GaussAppell_normalised}
{}_2\cF_1(\alpha,\beta_1,\gamma;y_1) &\,= \cF_D^{(1)}(\alpha,\beta_1,\gamma;y_1) = \beta(\gamma-\alpha, \alpha)\,{}_2F_1(\alpha,\beta_1,\gamma;y_1) \,,\\
\nonumber
\cF_1(\alpha,\beta_1,\beta_2,\gamma;y_1,y_2) &\,= \cF_D^{(2)}(\alpha,\beta_1,\beta_2,\gamma;y_1,y_2)
=\beta(\gamma-\alpha, \alpha)\,F_1(\alpha,\beta_1,\beta_2,\gamma;y_1,y_2) \,.
\end{align}

The Lauricella functions $F_D^{(r)}$ have been discussed in detail in refs.~\cite{brown_lauricella_2019} in the framework of twisted cohomology groups. In the following we briefly review the structure of the relevant twisted (co-)homology groups. From the integrand in eq.~\eqref{EQU.108}, we can read off the twist:
\begin{align}
\Phi=
 x^{a} (1-x)^{c-a}
 \prod_{j=1}^{r} (1-y_{j} x)^{-b_j}\, , 
\end{align} 
where $a,b_j,c$ are the non-integer parts of $\alpha, \beta_j,\gamma$. 
For the corresponding $(r+1)$-dimensional twisted (co-)homology groups $H_{\text{dR}}^1(\mathbb{C}-\Sigma_r, \nablasub{\underline{\mu}})$  and $H_1(\mathbb{C}-\Sigma_r, \check{\mathcal{L}}_{\underline{\mu}})$ (with $\underline{\mu} = (a,c-a,-\underline{b})$)
one can choose the bases 
\begin{align}
\label{bascocyLau}
\varphi_{r,j}= \rd\!\log\left(\frac{x}{x-y_{j-1}^{-1}}\right)\text{ for } 1\leq j \leq r+1 
\end{align} 
  and
\begin{align}
\label{laurcyc.1}
\gamma_{r,j} =\eta_{j-1,c}\otimes \Phi|_{\eta_{j-1,c}}\,,
    \end{align}
  where $\eta_{j-1,c}$ denotes the regularised version of the open interval $\eta_{j-1}=(0,y_{j-1}^{-1})$ (see appendix~\ref{sec.hom}) and $y_0=1$. Then:
\begin{align}
\label{eq:Lauricella1}
\mathcal{F}_D^{(r)}(a, \underline{b},c;\underline{y})=  \langle \varphi_{r,1} |\gamma_{r,1}]\, . 
\end{align}
All entries of the period matrix are  Lauricella functions and can be found in appendix~\ref{app.perlaur} together with the cohomology and homology intersection matrices. 

\subsubsection{Generalised hypergeometric ${}_{p+1}F_p$ functions} 
\label{sec:pFq}

The generalised hypergeometric ${}_{p+1}F_p$ functions are defined by the series 
\begin{align}
    {}_{p+1}F_p \left(\alpha_0,\alpha_1,\dots, \alpha_p , \beta_1, \dots, \beta_p, y\right) = \sum_{n=0}^\infty \frac{(\alpha_0)_n\dots (\alpha_p)_n}{(\beta_1)_n\dots(\beta_p)_n} \frac{y^n}{n!} \, , 
\end{align}
which converges on $|y|<1$, and the $\beta_i$ are not allowed to be negative integers. Additionally we assume that $\alpha_i-\beta_j\notin \mathbb{Z}$ and $\beta_i-\beta_j\notin \mathbb{Z}$ for any $i,j$. For $p=1$, we recover the definition of Gauss' hypergeometric function ${}_2F_1$. The generalised hypergeometric functions admit the integral representation:
\beq\bsp
    \label{fpp1int}
  & {}_{p+1}F_p \left(\alpha_0, \alpha_1,\dots, \alpha_p , \beta_1, \dots, \beta_p, y\right) =
\beta(\alpha_p,\beta_p-\alpha_p)^{-1}\, \\
 &  \times \int_0^1\rd t_p\, t_p^{\alpha_p-1}(1-t_p)^{\beta_p-\alpha_p-1} {}_{p}F_{p-1}(\alpha_0, \dots, \alpha_{p-1}, \beta_1,\dots, \beta_{p-1}, y  t_p) \\
 &= \left(\prod_{i=1}^p\beta(\alpha_i,\beta_i-\alpha_i)^{-1} \int_0^1\rd t_i\, t_i^{\alpha_i-1}(1-t_i)^{\beta_i-\alpha_i-1}\right)\,(1-yt_1\ldots t_p)^{-\alpha_0}\,,
\esp\eeq
with ${}_1F_0(\alpha_0,y)=(1-y)^{-\alpha_0}$. We also define the normalised version
\beq\bsp
\label{eq:pFq_normalised}
   {}_{p+1}\cF_p \left(\underline{\alpha}, \underline{\beta}, y\right) &\, = {}_{p+1}F_p \left(\underline{\alpha}, \underline{\beta}, y\right)\,\prod_{i=1}^p\beta(\alpha_i,\beta_i-\alpha_i)\\
 &\,= \left(\prod_{i=1}^p \int_0^1\rd t_i\, t_i^{\alpha_i-1}(1-t_i)^{\beta_i-\alpha_i-1}\right)\,(1-yt_1\ldots t_p)^{-\alpha_0}\,.
\esp\eeq
where we set $\underline{\alpha} = (\alpha_0, \alpha_1,\dots, \alpha_p)$ and $\underline{\beta} = (\beta_1, \dots, \beta_p)$. At this point we have to make a comment. We see that the zeroes of the twist correspond to the hyperplanes $t_i=0$ and $t_i=1$, as well as the hypersurface $1-yt_1\ldots t_p$. For $p\ge 1$, this hypersurface is not a hyperplane, so it appears that the generalised hypergeometric functions ${}_{p+1}F_p$ do not fall into the class of Aomoto-Gelfand hypergeometric functions. We can, however, change variables according to $t_i = \frac{x_{i}}{x_{i-1}}$ (with $x_0=1$), and we have
\beq
\label{eq:pFq_change_of_vars}
1-t_i = \frac{x_{i-1}-x_{i}}{x_{i-1}} \textrm{~~~and~~~} 1-yt_1\ldots t_p = 1-x_py\,.
\eeq
Hence, we see that in these variables, the zeroes of the twist are located on the hyperplanes $x_i=0$, $x_i=x_{i-1}$ and $1-yx_p=0$, and so the generalised hypergeometric functions ${}_{p+1}F_p$ are indeed Aomoto-Gelfand hypergeometric functions.

The twisted cohomology theory relevant for the generalised ${}_{p+1}F_p$ function was discussed in refs.~\cite{mimachi_intersection_2010,goto2014intersection}.  
In the following, we review the case of the ${}_3F_2$ function, $p=2$, following ref.~\cite{goto2014intersection}. 
The cases with $p>2$ are similar, and we refer to appendix~\ref{perfp}.

\begin{figure}[!t]
\begin{center}
\includegraphics[scale=0.25]{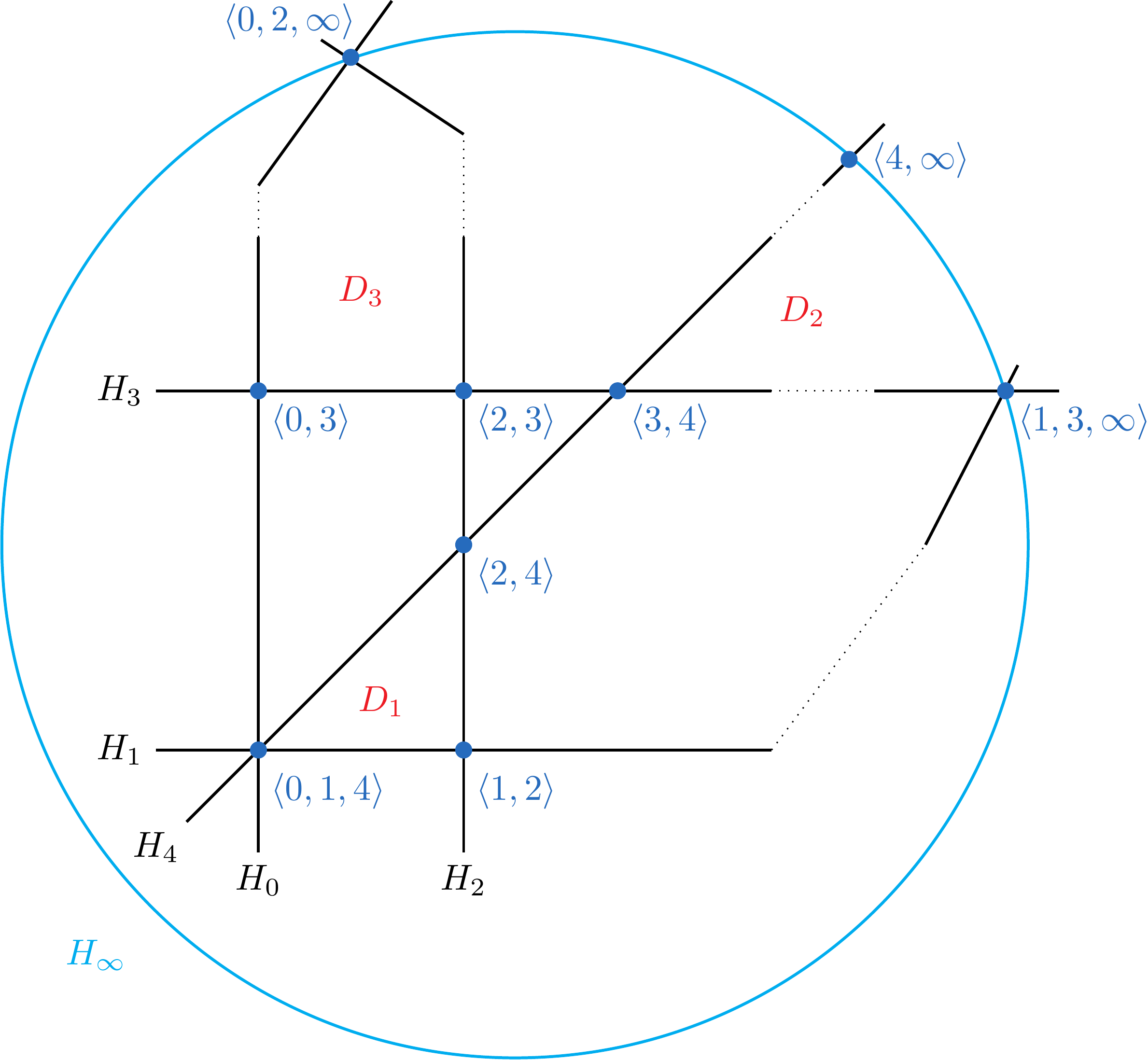}
\end{center}
\caption{\label{fig.f32space} A projection of $X_{{}_3F_2}$ with $y<1$ to $\mathbb{R}^2$. The hyperplanes of $\Sigma$ are represented by $H_0=\{x_1=0\},H_1=\{x_2=0\},H_2=\{x_1=1\},H_3=\{x_2=y^{-1}\}$ and $H_4=\{x_1=x_2\}$}
\end{figure}

For $p=2$, eq.~\eqref{eq:pFq_normalised} reduces to 
\begin{align}    \label{3f2int}
   {}_{3}\cF_2& \left(\underline{\alpha}, \underline{\beta}; y\right)\\
\nonumber   & = \int_0^1\int_0^1 t_2^{\alpha_2-1} (1-t_2)^{\beta_2-\alpha_2-1} t_1^{\alpha_1-1} (1-t_1)^{\beta_1-\alpha_1-1} (1-y t_1 t_2)^{-\alpha_0} \rd t_1\wedge \rd t_2\\
\nonumber   &=\int_{D_1} x_1^{\alpha_1-\beta_2} x_2^{\alpha_2-1} (1-x_1)^{\beta_1-\alpha_1-1} (1-yx_2)^{-\alpha_0} (x_1-x_2)^{\beta_2-a_2-1} \rd x_1\wedge \rd x_2\,,
\end{align}
where the last step follows from the change of variables in eq.~\eqref{eq:pFq_change_of_vars}, and the integration region $D_{1}$ will be defined below.
From this integral representation we can identify the twist:
\begin{align}
\label{twistf32}
    \Phi=  x_1^{a_1-b_2} x_2^{a_2} (1-x_1)^{b_1-a_1} (1-y x_2)^{-a_0} (x_1-x_2)^{b_2-a_2} \, , 
\end{align}
where $a_i, b_i$ are the non-integer parts of $\alpha_i,\beta_i$, and $\underline{a} = (a_1-b_2,a_2,b_1-a_1,-a_0,b_2-a_2)$. 
The zeroes of the twist are located at 
\beq
    \Sigma_{{}_3F_2}= \{ x_1=0\}\cup\{x_2=0\}\cup \{x_1=1\}\cup \{x_2=y^{-1}\}\cup \{x_1=x_2\} \, . 
    \eeq 
  Let  $X_{{}_3F_2}=\mathbb{C}^2-\Sigma_{{}_3F_2}$. A projection of $X_{{}_3F_2}$to $\mathbb{R}^2$  with $y<1$  is depicted in figure \ref{fig.f32space}.
For the basis of $H_{\text{dR}}^2(X_{{}_3F_2}, \nablasub{\underline{a}})$ we choose the co-cycles 
\beq\bsp
    \label{cohobas3f2}
    \varphi_{1} &\,=\frac{\rd x_2\wedge \rd x_2}{x_2 (1-x_1)(x_1-x_2)}\, ,\\
    \varphi_{2}&\,=\frac{y\, \rd x_1\wedge \rd x_2}{(1-y x_2)(x_1-x_2)}\,,\\
    \varphi_{3} &\,=\frac{y\, \rd x_1\wedge \rd x_2}{x_1(1-x_1)(1-y x_2)}\, . 
\esp\eeq
The basis elements $\gamma_i$ of the corresponding twisted homology group can be chosen to be supported on the following chambers (see figure~\ref{fig.f32space}):
\beq\bsp
\label{subf32}
    D_{1}&=\{0<x_2<x_1<1\}\,,\\
    D_{2}&=\{y^{-1}<x_2<x_1<\infty\}\,,\\
    D_{3}&=\{0<x_1<1, y^{-1}<x_2<\infty\} \, . 
\esp\eeq
We can see that with these definitions we have:
\beq
{}_3\mathcal{F}_2(a_0,a_1,a_2,b_1,b_2 ,y) =  \langle \varphi_{1}|\gamma_{1}]\,.
\eeq
All period and intersection matrices for this basis are given in appendix~\ref{perfp}.


\subsection{Single-valued analogues of hypergeometric functions}
\label{subsec.svpm}

In general, hypergeometric functions are multi-valued functions in the entries of $\bfY$. The multi-valuedness and the monodromies are most easily understood for the period matrix. If we analytically continue the period matrix along some closed contour $\delta$ in the space of parameters $\bfY_{ij}$, then after analytic continuation the period matrix must still satisfy the differential equation~\eqref{eq:GM}. As a consequence, $\bP_{\!\underline{a}}$ transforms under analytic continuation as 
\begin{align}
\label{EQU.26}
\bP_{\!\underline{a}} \rightarrow \bP_{\!\underline{a}}\, \bM_{\underline{a}}({\delta})\, .  
\end{align}
For hypergeometric functions, the monodromy matrix $\bM_{\underline{a}}({\delta})$  is generally a rational function of phase factors $e^{2\pi ia_k}$. From now on we assume that the entries of $\underline{a}$ are real. However, as already pointed out in ref.~\cite{brown_lauricella_2019}, the final results do not crucially depend on this assumption. It is easy to see that, if $\underline{a}$ is real, we have:
\begin{align}
\label{EQU.27}
\overline{\bM}_{-\underline{a}}(\delta) =\bM_{\underline{a}}(\delta)\, , 
\end{align}
where the bar denotes complex conjugation. In the remainder of this section we will define a class of functions that satisfy the same \emph{holomorphic} differential equation~\eqref{eq:GM_holom} as the period matrix, but are single-valued, i.e., they transform trivially under analytic continuation. In the following we start by motivating the mathematical construction without giving mathematical proofs. For the proofs we refer to the mathematical literature, e.g.,~refs.~\cite{Hanamura1,brown_lauricella_2019,Brown:2019wna,Brown:2018omk}.

We define a matrix:
\begin{align}
\label{EQU.28}
\bP^{\text{sv}}_{\!\underline{a}} = {\bP}_{\!\underline{a}} \overline{\bP}_{\!-\underline{a}}^{-1} \overline{\bC}_{-\underline{a}} = (-2\pi i)^{-n}\,\bP_{\!\underline{a}}\,\big({\bH}_{\underline{a}}^{-1}\big)^{T}\, \overline{\bP}_{\!\underline{a}}^T \,,
\end{align}
where the last step follows from the twisted Riemann bilinear relations in eq.~\eqref{EQU.24}, and we used the fact that intersection numbers between twisted cycles are rational functions with rational coefficients of $e^{2\pi ia_j}$, which implies that $\overline{\bH}_{-\underline{a}} = {\bH}_{\underline{a}}$.
It is easy to see that the matrix $\bP^{\text{sv}}_{\!\underline{a}}$ satisfies the same holomorphic differential equation~\eqref{eq:GM_holom} as the period matrix ${\bP}_{\!\underline{a}}$. Indeed, since ${\bP}_{\!\underline{a}}$ is the only holomorphic factor in eq.~\eqref{EQU.28}, we have
\begin{align}
\label{EQU.30}
\partial\bP^{\text{sv}}_{\!\underline{a}} = \left(\partial {\bP}_{\!\underline{a}}\right) \overline{\bP}_{-\underline{a}}^{-1}\overline{\bC}_{-\underline{a}} = \bA_{\underline{a}}\bP^{\text{sv}}_{\!\underline{a}} \, .
\end{align}
Under monodromy transformations as in eq.~\eqref{EQU.26}, this particular solution is single-valued. To see this, note that intersection numbers between twisted co-cycles are rational functions, and so they are single-valued. Then, using eqs.~\eqref{EQU.26} and~\eqref{EQU.27}, we have:
\begin{align}
\label{EQU.29}
\bP^{\text{sv}}_{\!\underline{a}} \rightarrow {\bP}_{\!\underline{a}}\, {\bM}_{\underline{a}}(\delta) \overline{\bM}_{-\underline{a}}(\delta)^{-1} \overline{\bP}_{\!-\underline{a}}^{-1} \overline{\bC}_{-\underline{a}}= \bP^\text{sv}_{\!\underline{a}}\, . 
\end{align} 
To summarise, we see that the entries of $\bP^{\text{sv}}_{\!\underline{a}}$ satisfy the same holomorphic differential equation as the entries of $\bP_{\!\underline{a}}$, but they are single-valued. Since the entries of $\bP_{\!\underline{a}}$ form a basis for a given class of hypergeometric function (with a given twist), it is natural to call the entries of $\bP^{\text{sv}}_{\!\underline{a}}$ the \emph{single-valued analogues} of these hypergeometric functions, and we call $\bP^{\text{sv}}_{\!\underline{a}}$ the \emph{single-valued period matrix}\footnote{Note that there is an ambiguity in this naming: the object $\bP_{\!\underline{a}}\overline{\bP}^{-1}_{\!-\underline{a}}$ itself is also commonly called the single-valued period matrix in the literature, cf.~refs.~\cite{brown_lauricella_2019,Tapuskovic:2023xiu}. In our context, it is more natural to absorb the matrix $\overline{\bC}_{-\underline{a}}$ into the definition of the single-valued period matrix, because only the combination in eq.~\eqref{EQU.29} shows up.}.
The entries of $\bP^{\text{sv}}_{\!\underline{a}}$ admit the integral representation~\cite{brown_lauricella_2019,Brown:2019wna,Brown:2018omk}:
\beq
\big(\bP^{\text{sv}}_{\!\underline{a}}\big)_{ij} =\left(-\frac{1}{2\pi i }\right)^n \int_X |\Phi|^2\,  {\varphi}_i\wedge \overline{\varphi}_j\, . 
\eeq
More generally, to every pair of twisted co-cycles $\varphi_A$ and ${\varphi}_B$ we can associate their single-valued  period:
\beq
\left(-\frac{1}{2\pi i }\right)^n \int_{X} |\Phi|^2\,  {\varphi_A}\wedge \overline{\varphi}_B\,.
\eeq

We can now apply this contruction to the particular case of the Aomoto-Gelfand hypergeometric functions and define single-valued analogues for them. We define (cf.~eq.~\eqref{EQU.107}):
\begin{align}
\label{EQU.104} 
\cF^{\text{sv}}(\underline{\alpha},{\bfY}) = \left(-\frac{1}{2\pi i}\right)^n\,\int_{\mathbb{C}^n}
|\Phi|^2\,\omega\wedge \overline{\omega}\, , 
\end{align} 
where $\omega_n$ was defined in eq.~\eqref{eq:topform}. We see that in this way we can associate to every class of holomorphic hypergeometric functions with basis given by the period matrix $\bP_{\!\underline{a}}$ a class of non-holomorphic hypergeometric functions with basis $\bP_{\!\underline{a}}^{\text{sv}}$ that satisfies the same holomorphic differential equation, but it is single-valued. Evaluating the integral in eq.~\eqref{EQU.104} may be a non-trivial task. However, we can use eq.~\eqref{EQU.28} to relate this integral to a bilinear in ordinary hypergeometric functions. For example, consider a hypergeometric function $\cF(\underline{a},\bfY)$ as in eq.~\eqref{EQU.107} with all $a_i$ non-integer for simplicity, and assume that we have chosen a basis of twisted cycles and co-cycles such that $(\bP_{\!\underline{a}})_{11}=\cF(\underline{a},\bfY)$. Then the function $\cF^{\text{sv}}(\underline{a},{\bfY})$ from eq.~\eqref{EQU.104} is given by
\beq\label{eq:cFsv_bilinear}
\cF^{\text{sv}}(\underline{a},{\bfY}) = (\bP_{\!\underline{a}}^{\text{sv}})_{11} = \left(-\frac{1}{2\pi i}\right)^n\,(\bP_{\!\underline{a}})_{1i} \,(\bH_{\underline{a}}^{-1})_{ji}\,(\overline{\bP}_{\!\underline{a}})_{1j}\,.
\eeq
We see that, in this particular basis, $\cF^{\text{sv}}(\underline{a},{\bfY})$ is determined once we know the matrix of intersection numbers $\bH_{\underline{a}}$ and the first row of the period matrix $\bP_{\!\underline{a}}$. The computation of the relevant intersection numbers is known in the literature, and reviewed in appendix \ref{sec.hom}. We will give examples of the single-valued analogues of the hypergeometric functions from the previous sections. In these examples we take $\alpha_i=a_i<1$ to be non-integer for simplicity and because we only consider Feynman integrals with all propagator weights non-integer in the following section.

\paragraph{The single-valued Beta function.}
Let us start the discussion by reviewing the single-valued analogue of Euler's Beta function from section~\ref{sec:beta_example}.
From eq.~\eqref{EQU.104}, we see that it admits the integral representation:
\begin{align}
\label{EQU.89}
\beta^{\text{sv}} (a_1,a_2)= - \frac{1}{2\pi i }\int_{\mathbb{C}} |z|^{2a_1} |1-z|^{2a_1} \frac{\rd z \wedge \rd \bar{z}}{|z|^2|1-z|^2}  \, . 
\end{align}
An expression for $\beta^{\text{sv}}$ in terms of Gamma functions can be obtained from the period and intersection matrix in eqs.~(\ref{EQU.82}) and (\ref{EQU.83}): 
\begin{align}
 \label{EQU.84A}
\beta^{\text{sv}} (a_1,a_2)
=
- \frac{1}{2\pi i }\,\bP_{\!\underline{a}}\bH_{\underline{a}}^{-1} \overline{\bP}_{\!\underline{a}}
 =\frac{\Gamma(a_1)\Gamma(a_2)\Gamma(1-a_1-a_2)}{\Gamma(a_1+a_2)\Gamma(1-a_1)\Gamma(1-a_2)}\, , \end{align}
 where we dropped the transposition of all matrices, because they are one-dimensional.
This was derived in ref.~\cite{Brown:2019wna,Brown:2018omk} but already appeared in refs.~\cite{beta}.


\paragraph{Single-valued Lauricella functions.} 
\label{secsvlau}

The single-valued analogues of the Lauricella $F_D$ functions have been considered in detail in ref.~\cite{brown_lauricella_2019}. They admit the integral representation (we follow the notations of section~\ref{sec:Lauricella}):
\beq\bsp
\label{gensvlaur}
&\mathcal{F}_{D}^{\text{sv}}(a,\underline{b}, c;\underline{y})  =-\frac{1}{2\pi i}\int_{\mathbb{C}} \rd x\wedge \rd \bar{x} \,|x|^{2(a-1)}|1-x|^{2(c-a-1)}\,\prod_{i=1}^{r}|1-y_ix|^{-2b_i} \, .
\esp\eeq
We can write $\mathcal{F}_{D}^{\text{sv}}(a,\underline{b}, c;\underline{y})$ as a bilinear in holomorphic Lauricella functions using eq.~\eqref{eq:cFsv_bilinear}. The expressions for the relevant period and intersection matrices can be found in eqs.~(\ref{intGenLa}) and~(\ref{laurper}). Here we only present the formulas for the special cases $r=1$ and $2$, which correspond to Gauss' hypergeometric function ${}_2F_1$ and the Appell $F_1$ function (cf.~eq.~\eqref{eq:GaussAppell}). 

For the single-valued analogue of Gauss' hypergeometric function, we find
\beq\bsp
\label{EQU.79}
{}_2\mathcal{F}_1^{\text{sv}} (a,b_1,c;y)=&\frac{\mathfrak{s}( a) \mathfrak{s}(c-a)}{\pi \mathfrak{s}( c)} {}_2\mathcal{F}_1(a,b_1,c;y){}_2\mathcal{F}_1(a,b_1,c;\bar{y})  
\\
& -\frac{\mathfrak{s}( b_1)\mathfrak{s}( c-b_1)}{\pi \mathfrak{s}( c)} {}_2\mathcal{G}_1(a,b_1,c;y){}_2\mathcal{G}_1(a,b_1,c;\overline{y}) \,,
\esp\eeq
where we introduced the abbreviation $\mathfrak{s}( x) = \sin(\pi x)$. The function ${}_2\mathcal{F}_1(a,b_1,c;{y}) $ is the normalised Gauss hypergeometric function introduced in eq.~\eqref{eq:GaussAppell_normalised}, and the function ${}_2\mathcal{G}_1(a,b_1,c;y)$ is given by
\beq
{}_2\mathcal{G}_1(a,b_1,c;y) = (-1)^{c-a-b_1} y^{1-c} {}_2\mathcal{F}_1\left( 1+a-c,1+b_1-c, 2-c;y\right)\, . 
\eeq
This expression agrees with the one given in ref.~\cite{brown_lauricella_2019}.





The single-valued Appell $F_1$ given by
\begin{align}
\label{F11DIntSV}
&\mathcal{F}_1^{\text{sv}}(a,b_1,b_2,c;y_1,y_2)=y_1^{-a } y_2^{-a }\frac{ \mathfrak{s}(a)\mathfrak{s}(b_1 ) \mathfrak{s}(b_2) \mathfrak{s}  ( a-c ) }{\mathfrak{s}  (b_1 +b_2-c )}\\
&\Bigg\{2\mathcal{F}_1(a,b_1,b_2,c;\bar{y}_1,\bar{y}_2)\Bigg[\frac{y_1^a y_2^a}{i}\frac{\mathfrak{s}(a-b_1-b_2)}{\mathfrak{s}(a)\mathfrak{s}(b_1)\mathfrak{s}(b_2)}\mathcal{F}_1(a,b_1,b_2,c;y_1,y_2)\notag\\
&\, \, \, \, \, \, \, \, \, \, \, \, +y_1^a (i+\mathfrak{c}(a))(1-i\mathfrak{c}(b_1))\mathcal{F}_1\left(a,1+a-c, b_1,1+a-b_2;\frac{1}{y_2},\frac{y_1}{y_2} \right)\notag \\
&\, \, \, \, \, \, \, \, \, \, \, \, - y_2^a (i+\mathfrak{c}(a))(1+i\mathfrak{c}(b_2)) \mathcal{F}_1\left(a, 1+a-c,b_2,1+a-b_1;\frac{1}{y_1},\frac{y_2}{y_1}\right) \Bigg]\notag\\
&-\frac{e^{-i \pi (2(a+b_1)+c)} y_2^{a}\bar{y}_2^{-a}  }{\mathfrak{s}(a)\mathfrak{s}(a-c)}\mathcal{F}_1\left(a, 1+a-c, b_1,1+a-b_2;\frac{1}{\bar{y}_2},\frac{\bar{y}_1}{\bar{y}_2}\right)\notag \\
&\, \, \, \, \, \times \Bigg[\frac{y_1^{a}\left(e^{i\pi(2a+b_1)}-e^{i\pi(b_1+2c)}\right)}{\mathfrak{s}(b_1)} \mathcal{F}_1(a,b_1,b_2,c;y_1,y_2)\notag\\
&\, \, \, \, \, \, \, \, \, \, \, \, +2 i e^{2i \pi (a+b_1)} \mathcal{F}_1\left(a,1+a-c, b_2, 1+a-b_1;\frac{1}{y_1},\frac{y_2}{y_1}\right)\Bigg]\notag\\
&-2i \frac{e^{-i \pi (a-b_2)}y_1^{a}\bar{y}_1^{-a}}{\mathfrak{s}(a)} ´\mathcal{F}_1\left(a, 1+a-c, b_2,1+a-b_1;\frac{1}{\bar{y}_1},\frac{\bar{y}_2}{\bar{y}_1}\right) \notag\\
&\, \, \, \, \, \times\Bigg[\frac{y_2^{a}}{\mathfrak{s}(b_2)} \mathcal{F}_1(a,b_1,b_2,c;y_1,y_2) + \frac{e^{i\pi (a-b_2+c)}}{\mathfrak{s}(a-c)} \mathcal{F}_1\left(a,1+a-c, b_1,1+a-b_2;\frac{1}{y_2},\frac{y_1}{y_2}\right) \Bigg]\notag
\\
& +\Bigg[\frac{2i y_1^a \bar{y}_2^{-a} \mathfrak{s}(b_1-c)}{\mathfrak{s}(a)\mathfrak{s}(b_1)\mathfrak{s}(a-c)}\Big|\mathcal{F}_1\left(a,1+a-c, b_1,1+a-b_2;\frac{1}{y_2},\frac{y_1}{y_2}\right)\Big|^2 +(y_1,b_1\leftrightarrow y_2,b_2)\Bigg]\Bigg\}\,,\notag
\end{align}
with $\mathfrak{c}(x)=\cot(\pi x)$. For the construction of the analytic continuations of the Appell $F_1$ function necessary to evaluate this function, we refer to ref.~\cite{bezrodnykh_analytic_2017}.

\paragraph{Single-valued versions of generalized hypergeometric ${}_{p+1}F_p$ functions.}  
\label{sec.homf32}

The intersection theory of the generalized hypergeometric ${}_{p+1}F_p$ functions has been considered in refs.~\cite{mimachi_intersection_2010,goto2014intersection}. Its single-valued analogoue is given by the integral (we use the conventions of section~\ref{sec:pFq}):
\beq\bsp
    \label{fpp1intsv}
  & {}_{p+1}\mathcal{F}_p^{\text{sv}} \left(\underline{a},\underline{b};y\right) =\left(-\frac{1}{2\pi i}\right)^p\int_{\mathbb{C}^p}\rd x_1\wedge\dots\wedge\rd x_p\wedge \rd \bar{x}_1\wedge\dots\wedge\rd\bar{x}_p\,|x_p|^{2(a_p-1)}\\
 &\times |1-x_1|^{2(b_1-a_1-1)} |1-yx_p|^{-2a_0}\,\prod_{k=1}^{p-1} |x_k|^{2(a_k-b_{k+1})} |x_k-x_{k+1}|^{2(b_{k+1}-a_{k+1}-1)} \, , 
\esp\eeq
which is equal to the following sum of bilinears of ${}_{p+1}\mathcal{F}_p$ functions: 
\begin{align}
    \label{svfm}
  &{}_{p+1}\mathcal{F}_p^{\text{sv}}\left(\underline{a},\underline{b};y\right) = 
    \big|{}_{p+1}\mathcal{F}_p (\underline{a},\underline{b};y)\big|^2\,\prod_{i=1}^p\frac{\mathfrak{s}(a_i)\mathfrak{s}(b_i-a_i) }{\pi \mathfrak{s}(b_i)}\\
        &-\sum_{i=1}^p |y|^{2(1-b_i)}\,\big| {}_{p+1}\mathcal{F}_p(\underline{a}_i,\underline{b}_i;y )\big|^2\, \frac{\mathfrak{s}(a_0)\mathfrak{s}(b_i-
        a_0)}{\pi\mathfrak{s}(b_i)} \prod_{j=1,j\neq i}^p \frac{\mathfrak{s}(b_i-a_j)\mathfrak{s}(b_j-a_j)}{\pi\mathfrak{s}(b_i-b_j)} \, ,\notag
\end{align}
where we defined
\beq\bsp
\underline{a}_i&\,=(1+a_i-b_i, 1+a_1-b_i,\dots,\widehat{1+a_i-b_i},\ldots,1+a_p-b_i,1+a_0-b_i)\,,\\
\underline{b}_i&\,=(1+b_1-b_i,\ldots,\widehat{1+b_i-b_i}, \ldots1+b_p-b_i,2-b_i)\,.
\esp\eeq
\section{Feynman integrals from single-valued twisted periods} 
\label{SVPF}

\label{subsec.svpm1d}

\subsection{The main theorem}

In this section we present our main result, namely we explain how Feynman integrals with massless propagators in two dimensions are related to single-valued analogues of Aomoto-Gelfand hypergeometric functions. Our starting point is the representation of the Feynman integral given in eq.~\eqref{eq:2DIntegral}. The entire information on the Feynman graph $G$ is encoded in the differential $(L,0)$ form $\omega_G$ defined in eq.~\eqref{omegag}. The latter can be cast in the form
\beq
\label{eq:omegaG_PhiG}
\omega_G = \Phi_G\,\omega_L\,,
\eeq
where $\omega_L$ is defined in eq.~\eqref{eq:topform}, and we defined the twist:
\beq
\label{twistfeyn}
\Phi_G = \prod_{j=1}^E\frac{1}{L_{j}(\underline{x},\underline{y})^{\nu_j}}\,.
  \eeq
  Since the $L_j$ all define hyperplanes, we see that the differential form $\omega_G$ defines the integrand of an Aomoto-Gelfand hypergeometric function, with twist $\Phi_G$. The associated twisted cohomology group is $H_{\text{dR}}^L(X_G,\nablasub{-\underline{\nu}})$, where we defined
  \beq\label{eq:X_G_def}
  X_G = \mathbb{C}^L-\bigcup_{j}\big\{L_{j}(\underline{x},\underline{y})=0\big\}\,,
  \eeq
  and
  \beq
  \nablasub{-\underline{\nu}} = \rd + \rd\!\log\Phi_G\wedge \cdot= \rd - \sum_j\nu_j\,\rd\!\log L_{j}(\underline{x},\underline{y})\wedge \cdot\,.
  \eeq
 Equation~\eqref{eq:2DIntegral} then takes the form:
   \beq
  \label{eq:2DIntegral}
 \tI_G(\underline{\nu},\underline{y}) = (-1)^{\frac{L(L-1)}{2}} \left(-\frac{1}{2\pi i}\right)^L
 \int_{X_G} \big|\Phi_G\big|^2\,\omega_L\wedge \overline{\omega}_L
 \,.
 \eeq
 By comparing eqs.~\eqref{EQU.104} and~\eqref{eq:2DIntegral}, we see that $ \tI_G(\underline{\nu},\underline{y})$ is equal (up to an overall sign) to a single-valued analogue of an Aomoto-Gelfand hypergeometric function. We can summarise this result in the following way:
 \begin{quote}
 \begin{thm}
\label{twodtheo}
\emph{Every $L$-loop Feynman graph $G$ with massless propagators and non-integer propagator powers (including $\nu_{\infty}=-\sum_j\nu_j$) in $D=2$ dimensions determines  a twisted cohomology group} $H_{\text{dR}}^L(X_G,\nablasub{-\underline{\nu}})$\emph{. The value of }$\tI_G$ \emph{is given by a single-valued analogue of an Aomoto-Gelfand hypergeometric function, which is a period of this cohomology group.}
\end{thm}
\end{quote}
This theorem describes the class of transcendental functions and their associated mathematical framework that arise from the computation of Feynman integral with massless propagators in two dimensions, both in momentum and in position space. There are not many cases where it is possible to characterise the class of functions that arise from Feynman integrals in such generality. 

We have seen in section~\ref{subsec.svpm} that every single-valued twisted period can be written as a bilinear in holomorphic and anti-holomorphic twisted periods, cf.~eqs. \eqref{EQU.28} and \eqref{eq:cFsv_bilinear}. Theorem~\ref{twodtheo} therefore implies that every Feynman integral with massless propagators in two dimensions can be expressed as a bilinear in holomorphic and anti-holomorphic hypergeometric functions. In fact, this was observed in several special cases before. For example, in ref.~\cite{Mimachi_2003} it was shown that four-point functions in conformal field theories typically evaluate to single-valued analogues of hypergeometric functions. In ref.~\cite{Halder:2023nlp} it was observed that Feynman integrals arising from a certain conformal field theory in two dimensions are single-valued hypergeometric functions. In ref.~\cite{derkachov_basso-dixon_2019,Duhr:2022pch} fishnet integrals in two dimensions were expressed in terms of single-valued combinations of hypergeometric functions. In particular, in ref.~\cite{Duhr:2022pch}  these integrals were related to the K\"ahler potential or quantum volume of a family of Calabi-Yau varieties, which are single-valued. Our theorem also implies that the periods of these families of Calabi-Yau varieties are Aomoto-Gelfand hypergeometric functions, and it gives a constructive way to compute the relevant K\"ahler potential or quantum volume from the homology generators and their intersection numbers. Recently, in ref.~\cite{Derkachev:2022lay} the two-loop kite diagram in position space was computed in arbitrary dimension $D$, and, to the surprise of the authors of ref.~\cite{Derkachev:2022lay}, it was observed that for two dimensions the result is holomorphically separable. Our theorem unifies and extends all these observations from the literature. In particular, it shows that the holomorphic separability and the single-valuedness are very generic features of Feynman integrals with massless propagators in two dimensions. Notably, they are by no means restricted to Feynman integrals in conformal field theories.

In the remainder of this section we present concrete examples of Theorem~\ref{twodtheo}. We will identify infinite classes of Feynman integrals in two dimensions that evaluate the single-valued analogues of the Beta function, the Lauricella $F_D^{(r)}$ function and the generalised hypergeometric ${}_{p+1}F_p$ functions.

\subsection{Massless banana integrals}
\label{sec:bananas}
We start by considering the banana integrals in momentum space with massless propagators (see figure~\ref{fig:bananas}):
 \begin{figure}[!t]
\centering
\includegraphics[scale=0.5]{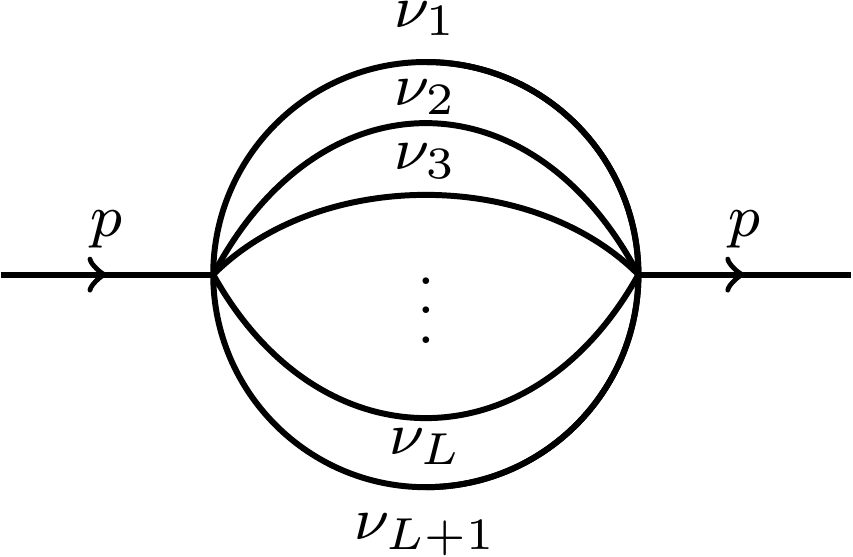}
\caption{The $L$-loop banana integral $\tI_{\text{Ban}_L}(\underline{\nu};p)$ in momentum space.}
\label{fig:bananas}
\end{figure}
\beq
\tI_{\text{Ban}_L}(\underline{\nu};p) = \left(-\frac{1}{2\pi i}\right)^L\int_{\mathbb{C}^L}\left(\bigwedge_{j=1}^L\rd k_j\wedge\rd \overline{k}_j\right) \prod_{j=1}^{L+1}\frac{1}{|k_{j-1}-k_j|^{2\nu_j}}\,,
\eeq
with $k_0=0$ and $k_{L+1}=p$. For $L=1$, we let $k_1=p\,x$, and by comparing to the definition of the single-valued Beta function in eq.~\eqref{EQU.89}, we find:
\beq
\tI_{\text{Ban}_1}(\underline{\nu};p) = |p|^{2-2\nu_{12}}\,\beta^{\text{sv}}(1-\nu_1,1-\nu_2)\,,
\eeq
with $\nu_{i_1\cdots i_k} = \nu_{i_1}+\ldots+\nu_{i_k}$.
Using eq.~\eqref{EQU.84A}, we can express this last equation as a ratio of Gamma functions. We stress that this result is not new, but it is the well-known expression for the one-loop bubble integral with arbitrary propagator powers. We nevertheless include it here, because this well-known result serves as an illustration of Theorem~\ref{twodtheo}. For $L>1$, we can recursively integrate the one-loop bubbles to obtain:
\beq
\tI_{\text{Ban}_L}(\underline{\nu};p) = |p|^{2L-2\nu_{1\cdots L}}\,\prod_{j=1}^L\beta^{\text{sv}}(j-\nu_{1\cdots j},1-\nu_j)\,.
\eeq
We see that all banana integrals with massless propagators in two dimensions can be written as a product of single-valued beta functions.

\subsection{One-loop integrals}
As a next example, let us consider a generic one-loop integral in two dimensions. We focus on the integral given by a star with $N$ external vertices in position space (see figure~\ref{fig:one-loop}):
 \begin{figure}[!t]
\centering
\includegraphics[scale=0.4]{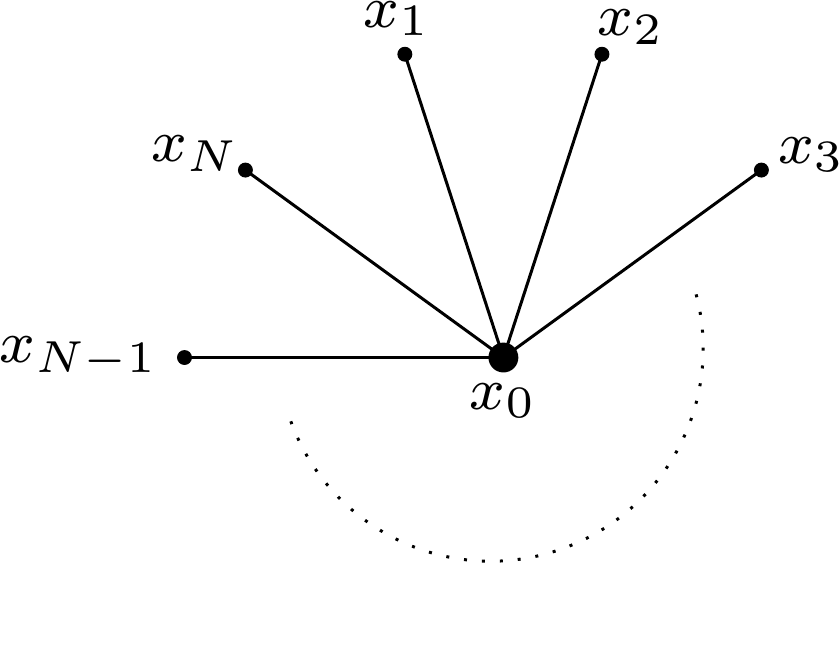}
\caption{The $N$-point star integral $\textrm{Star}_N(\underline{\nu};\underline{x})$ in position space.}
\label{fig:one-loop}
\end{figure}
\beq
\textrm{Star}_N(\underline{\nu};\underline{x}) = -\frac{1}{2\pi i}\int_{\mathbb{C}}\rd x_0\wedge \rd\overline{x}_0\,\prod_{j=1}^N\frac{1}{|x_{0j}|^{2\nu_j}}\,,
\eeq
with $x_{ij} = x_i-x_j$ and we defined $\underline{\nu}=(\nu_1,\ldots,\nu_N)$ and $\underline{x}=(x_1,\ldots,x_N)$. We note that this integral covers both one-loop integrals in position space and momentum space, because we can write the momentum space integral in terms of dual coordinates~\cite{Drummond:2006rz}. We obtain
\begin{align}
\textrm{Star}_N&(\underline{\nu};\underline{x}) =\\
\nonumber&=\,
\mathcal{F}_{D}^{\text{sv}}\big(1-\nu_1,\nu_3, \dots, \nu_N, 2-\nu_{12}; y_1, \dots, y_{N-2}\big)\,\frac{1}{|x_{12}|^{2\nu_{12}-2}}\,\prod_{j=3}^{N}\frac{1}{|x_{1j}|^{2\nu_j}}\,,
\end{align}
with $y_i = \frac{x_{12}}{x_{1(i+2)}}$.
We see that every one-loop integral can be expressed in terms of the single-valued analogue of Lauricella's $F_D^{(N-2)}$ function. The latter can be cast in the form of a bilinear in Lauricella $F_D^{(N-2)}$ functions via eq.~\eqref{F11DIntSV}. Note that, conversely, every $\mathcal{F}_{D}^{\text{sv}}$ function computes up to an overall factor the value of a one-loop Feynman integral in two dimensions.

For $N=2$ we recover the result for the one-loop bubble integral from section~\ref{sec:bananas}.
For $N=3,4$, in the case of the one-loop triangle and box integral, these integrals can be expressed in terms of Gauss' hypergeometric function ${}_2F_1$ or the Appell $F_1$ function, cf.~eqs.~\eqref{eq:GaussAppell} and~\eqref{eq:GaussAppell_normalised}:
\beq\bsp
\textrm{Star}_3(\underline{\nu};\underline{x}) &\,= |x_{12}|^{2-2\nu_{12}}\,|x_{13}|^{-2\nu_3}\,{}_2\mathcal{F}_{1}^{\text{sv}}\big(1-\nu_1,\nu_3, 2-\nu_{12}; y_1\big)\,,\\
\textrm{Star}_4(\underline{\nu};\underline{x}) &\,= |x_{12}|^{2-2\nu_{12}}\,|x_{13}|^{-2\nu_3}\,|x_{14}|^{-2\nu_4}\,\mathcal{F}_{1}^{\text{sv}}\big(1-\nu_1,\nu_3,\nu_4, 2-\nu_{12}; y_1,y_2\big)\,.
\esp\eeq
When $\sum_{j=1}^N\nu_j=2$ the integral transforms under the conformal algebra $\mathfrak{so}(1,3)$ in two Euclidean dimensions with conformal weight $\nu_j$ at each external point. In that case, the Lauricella $F_D^{(N-2)}$ functions that one may initially associate to the one-loop integral reduce to Lauricella $F_D^{(N-3)}$ functions. For example, consider the five-point function
\begin{align}
&\textrm{Star}_{5}(\underline{\nu},\underline{x}) \,= -\frac{1}{2\pi i}\,\int_{\mathbb{C}} \frac{ \rd x_0\wedge \rd\bar{x}_0}{\left( |x_{01}|^2|x_{02}|^2 |x_{04}|^2|x_{05}|^2\right)^{\frac{2}{3}}\,|x_{03}|^{\frac{4}{3}}}\,,
\end{align}
This integral arises in the computation of conformally invariant fishnet integrals with cubic vertices~\cite{fishnet_general}. If we multiply the integral by $|x_{13}|^{\frac{2}{3}}|x_{23}|^{\frac{2}{3}}|x_{45}|^{\frac{2}{3}}$, it becomes conformally invariant. We may then use a conformal transformation to fix $(x_3,x_4,x_5) = (\infty,0,1)$, and we obtain:
\begin{align}
\nonumber\textrm{Star}_{5}(\underline{\nu},\underline{x}) &\,= -\frac{1}{2\pi i\,|x_{13}|^{\frac{2}{3}}|x_{23}|^{\frac{2}{3}}|x_{45}|^{\frac{2}{3}}}\,\int_{\mathbb{C}} \frac{ \rd x_0\wedge \rd\bar{x}_0}{\left( |x_0|^2|1-x_0|^2 |x_0-u_1|^2|x_0-u_2|^2\right)^{\frac{2}{3}}}\\
&\,= \left|\frac{x_{13}\,x_{23}\,x_{45}^3}{x_{14}^2\,x_{24}^2\,x_{35}^4}\right|^{\frac{2}{3}}\,\cF_1^{\text{sv}}\left(\frac{1}{3},\frac{2}{3},\frac{2}{3},\frac{2}{3};u_1^{-1},u_2^{-1}\right)\\
\nonumber&\,= \frac{\sqrt{3}u_1^{\frac{1}{3}}u_2^{\frac{1}{3}} }{2}\left|\frac{x_{13}\,x_{23}\,x_{45}^3}{x_{14}^2\,x_{24}^2\,x_{35}^4}\right|^{\frac{2}{3}}\,\Bigg\{
2i\,\mathcal{F}_1\left(\frac{1}{3},\frac{2}{3},\frac{2}{3},\frac{2}{3}; \bar{u}_1^{-1},\bar{u}_2^{-1}\right)\\
\nonumber&\times\Bigg[
e^{-\frac{i\pi}{3}}\, u_2^{-\frac{1}{3}}\, \mathcal{F}_1\left(\frac{1}{3},\frac{2}{3},\frac{2}{3},\frac{2}{3}; u_1, \frac{u_1}{u_2}\right) - u_1^{-\frac{1}{3}}\, \mathcal{F}_1\left(\frac{1}{3},\frac{2}{3},\frac{2}{3},\frac{2}{3}; u_2,\frac{u_2}{u_1}\right)\Bigg]\\
\nonumber& +2e^{\frac{i\pi}{6}} \mathcal{F}_1\left(\frac{1}{3},\frac{2}{3},\frac{2}{3},\frac{2}{3};\bar{u}_1,\frac{\bar{u}_1}{\bar{u}_2}\right)\\
\nonumber &\times\Bigg[e^{\frac{2i\pi}{3}}\, u_2^{-\frac{1}{3}}\, \mathcal{F}_1\left(\frac{1}{3},\frac{2}{3},\frac{2}{3},\frac{2}{3};u_1^{-1},u_2^{-1}\right) + \mathcal{F}_2\left(\frac{1}{3},\frac{2}{3},\frac{2}{3},\frac{2}{3};u_2,\frac{u_2}{u_1}\right)\Bigg] \\
\nonumber & - 2i\,\mathcal{F}_1\left(\frac{1}{3},\frac{2}{3},\frac{2}{3},\frac{2}{3};u_2,\frac{u_2}{u_1}\right) \\
\nonumber & \Bigg[e^{\frac{i\pi}{3}}\,\mathcal{F}_1\left(\frac{1}{3},\frac{2}{3},\frac{2}{3},\frac{2}{3};u_1,\frac{u_1}{u_2}\right) - u_1^{-\frac{1}{3}}\, \mathcal{F}_1\left(\frac{1}{3},\frac{2}{3},\frac{2}{3},\frac{2}{3};u_1^{-1},u_2^{-1} \right) \Bigg]
\Bigg\}\,,
\end{align}
where we defined the cross ratios
\beq
u_i=\frac{(x_i-x_4)(x_3-x_5)}{(x_i-x_3)(x_4-x_5)}\,,
\eeq
and assume $u_1,u_2>0$.

\subsection{Ladder integrals} 
\label{oneex}

\begin{figure}[!t]
   \centering
    \includegraphics[scale=0.5]{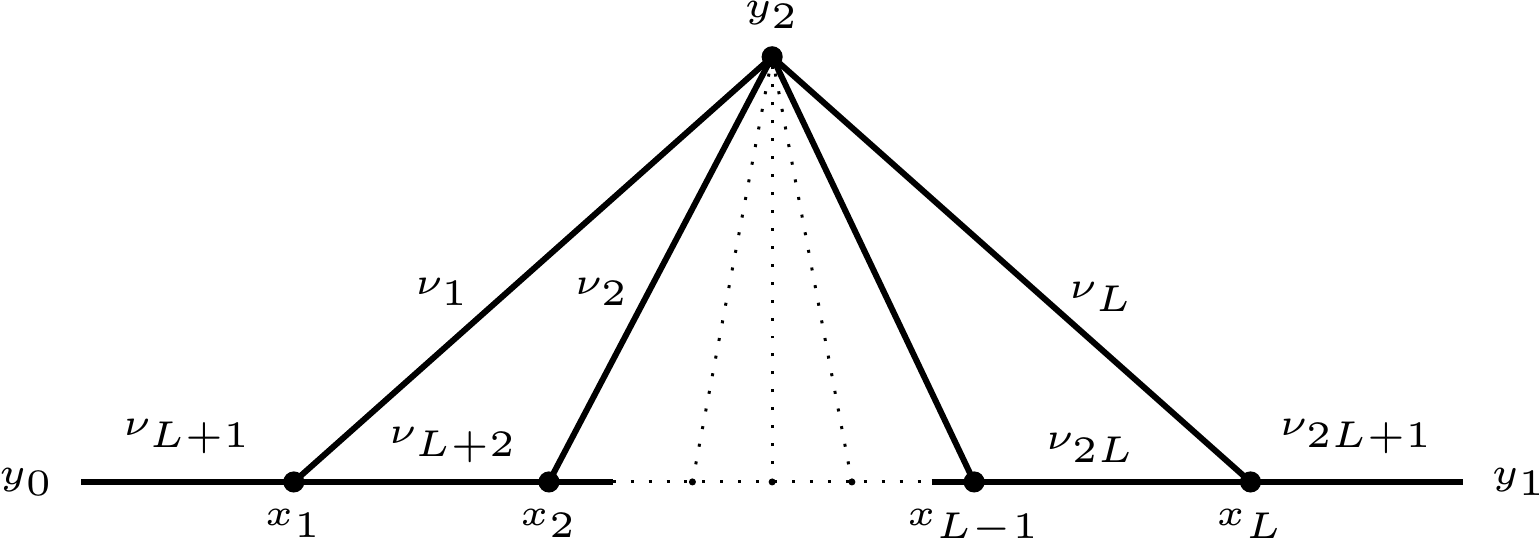}
    \caption{The $L$-loop ladder integral $\text{Lad}_L(\underline{\nu};\underline{y})$ in position space.}
    \label{genladfig}
\end{figure}

As a last example, we consider the $L$-loop ladder integrals in position space as shown in figure~\ref{genladfig}:
\begin{align}
\label{genlad11} 
 \text{Lad}_{L}& (\underline{\nu};y_0,y_1,y_2) = |y_0-y_2|^{2(2-\sum_{i=1}^{2L+1}\nu_i )}\,|y|^{2\nu_{2L+1}}\,\\
 &\,\times\left(-\frac{1}{2\pi i}\right)^{L}\int_{\mathbb{C}^{L}}\left(\bigwedge_{i=1}^L\rd x_i \wedge \rd \bar{x}_i\right)|1-y x_L|^{-2\nu_{2L+1}}\prod_{j=1}^L|x_j|^{-2\nu_j} |x_{j-1}-x_j|^{-2\nu_{L+j}}
\nonumber\, ,
\end{align}
where we defined
\beq
y = \frac{y_0-y_2}{y_1-y_2}\,.
\eeq
Note that for $L=1$ we recover the one-loop three-point function $\textrm{Star}_3$.
We can compare eq.~\eqref{genlad11} to eq.~\eqref{fpp1intsv}, and we see that
\begin{align}
\label{result1A}
 \text{Lad}_{L}& (\underline{\nu};y_0,y_1,y_2) =& (-1)^{\frac{L(L-1)}{2}}|y_0-y_2|^{2(2-\sum_{i=1}^{2L+1}\nu_i )}|y|^{2\nu_{2L+1}}{}_{L+1}\mathcal{F}_{L}^{\text{sv}}(\underline{a}, \underline{b}, y) \, ,
\end{align} 
where the $a_i$ and $b_i$ are given by:
\beq\bsp
\label{aitonui}
    a_0=\nu_{2L+1} \text{ and } a_i = L -i +1 -\sum_{j=i}^{2L} \nu_j + \sum_{j=L+1}^{L+i} \nu_j\text{ for } i >0\, , \\
    b_1=L+1-\sum_{j=1}^{2L} \nu_j \text{ and } 
    b_i= L-i+2-\sum_{j=i}^{2L} \nu_j+ \sum_{j=L+1}^{L+i-1}\nu_j \text{ for $i>1$}\, . 
\esp\eeq
A specific class of ladder integrals, which arise from conformal four-point functions, have been considered in ref.~\cite{derkachov_basso-dixon_2019}. In ref.~\cite{Derkachev:2022lay} the two-loop ladder integral has been evaluated in the limit $y_0=y_1$ in arbitrary dimension $D$, and we reproduce this result for $D=2$. In fact, it was observed that in $D=2$ the result can be written as a bilinear in ${}_3F_2$ functions.
Our result shows that this is not a coincidence, and a consequence of our general Theorem~\ref{twodtheo}, which for that graph specifies to 
\beq\bsp
 \text{Lad}_{2} &(\underline{\nu};y_1,y_2)= \\
 &= -|y_1-y_2|^{2(2-\sum_{i=1}^{5}\nu_i )}{}_{3}\mathcal{F}_{2}^{\text{sv}}(\nu_5, 2-\nu_{124},1-\nu_2,3-\nu_{1234}, 2-\nu_{24}, 1) \, .
\esp\eeq
Moreover, eq.~\eqref{result1A} generalises the results of refs.~\cite{derkachov_basso-dixon_2019,Derkachev:2022lay} to ladder integrals with an arbitrary number of loops and arbitrary propagator powers.
\section{Conclusion}

In this paper we have studied multi-loop Feynman integrals with massless propagators in two Euclidean dimensions. Our main result is that, at least as long as the exponents of the propagators are non integer, the integral is a single-valued analogue of an Aomoto-Gelfand hypergeometric function. The latter can themselves be expressed as bilinears in hypergeometric functions, using the double-copy formula for single-valued twisted periods obtained by Brown and Dupont in the context of string theory amplitudes~\cite{Brown:2019wna,Brown:2018omk}. Using this double-copy formula and algorithms to compute homology intersection numbers, it becomes possible to evaluate these Feynman integrals explicitly. We have done this in the case of the simplest and most well-known classes of hypergeometric functions, and we have shown that one-loop integrals correspond to the single-valued analogues of Lauricella $F_D^{(r)}$ introduced in ref.~\cite{brown_lauricella_2019}, while the generalised ${}_{L+1}F_L$ functions compute ladder integrals with $L$ loops. We find it remarkable that one can identify a well-defined class of functions that computes all Feynman integrals with massless propagators in two dimensions, and we are not aware of a similar result in other dimensions.

So far our result is restricted to the case where all propagator exponents are non integer. While this case is by itself interesting, because it covers, for example, correlation functions in two-dimensional CFTs, it would be interesting to understand how this extends to integer exponents. While in many applications it is possible to take appropriate limits, mathematically the situation of integer exponents is very different from the twisted cohomology setting considered here. If some, but not all, exponents are integers, the correct setting is \emph{relative} twisted cohomology~\cite{matsumoto2019relative,Caron-Huot:2021iev}. If instead all exponents are integer, the twist is trivial, and twisted cohomology is not applicable anymore at all. We leave this for future work.
\subsection*{Acknowledgments}
We thank Florian Loebbert, Albrecht Klemm and Christoph Nega for discussions and collaboration on a related project. FP thanks Andrzej Pokraka, Carlos Rodriguez, Oliver Schlotterer and  Cathrin Semper for discussions and Robin Marzucca for collaboration on a related project. This work was co-funded by the European Union (ERC Consolidator Grant LoCoMotive 101043686). Views and opinions expressed are however those of the author(s) only and do not necessarily reflect those of the European Union or the European Research Council. Neither the European Union nor the granting authority can be held responsible for them.

\appendix

\section{Twisted homology groups and their intersection numbers}
\label{sec.hom}

An important tool to compute single-valued analogues of hypergeometric functions is the evaluation of the intersection numbers between twisted cycles. While the computation of intersection numbers between twisted co-cycles has been discussed extensively in the physics literature (cf.,~e.g.,~refs.refs.~\cite{Mizera:2017rqa,Mizera:2019ose,Frellesvig:2019uqt,Frellesvig:2020qot,Mizera:2019vvs}), intersection numbers between twisted co-cycles have not appeared that prominently in physics. One example, where the homology intersection numbers have a central role is in a construction of KLT relations in ref.~\cite{Britto:2021prf}.

In this appendix we review the computation of intersection numbers of twisted homology groups, assuming that the twist has only linear factors -- as is the case for the hypergeometric functions needed for Feynman integrals in two dimensions. The aim of this section is to give a pedagogical review of this topic, based  on refs.~\cite{MR0841131,aomotoold,aomoto_theory_2011,yoshida_hypergeometric_1997,matsubaraheo2023lectures,kita_intersection_1994-2,kita_intersection_2006-2,yoshida_intersection_2000-1,kita_intersection_2006-4,mimachi_intersection_2010,goto2020lauricellas,GOTO_2013,goto2014intersection}
We generally compute the intersection numbers in terms of the objects
\begin{align}
\label{23.4}
c_j = \exp(2\pi i a_j) \text{ and } d_j = c_j-1   \, , 
\end{align}
which naturally appear due to analytical continuations of the multi-valued twist. 
We denote their products by
\begin{align}
\label{23.5}
c_{jk\dots } = c_j c_k \dots \text{ and } d_{jk\dots} = c_{jk\dots} -1\, .
\end{align}
\subsection{Twisted cycles} 
\label{sec.hom.1} 

We work with the notation and conventions of section~\ref{sec.twist}, considering twisted cycles in a non-compact space $X=\mathbb{C}^n -\Sigma$ where $\Sigma$ is a collection of hyperplanes defined by linear equations specified by the singularities of the twist. 

Generally, we can take the locally-finite twisted cycles in eq.~(\ref{duatwistedcycle}) of $  H_{n}(X, {{\mathcal{L}}}_{\underline{a}}) $ to be supported on chambers bounded by the hyperplanes in $\Sigma$ and the twisted cycles to be regularised versions of these chambers, where the regularisation consists of taking $\epsilon$-tubular neighbourhoods around all bounding hyperplanes \cite{aomoto_theory_2011}. Note that one of the hyperplanes can also be the hyperplane at infinity. 
We discuss this in more detail for the case $n=1$. An example for $n=2$ can be found in the treatment of the hypergeometric ${}_3F_2$ function in appendix~\ref{perfp}.

\begin{figure}[!t]
\begin{center}\includegraphics[scale=0.3]{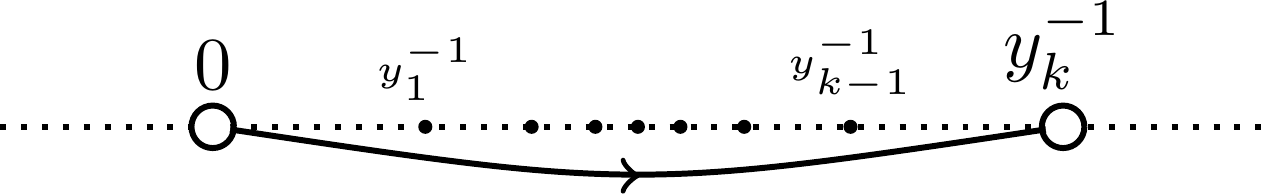}\, \, \, \includegraphics[scale=0.26]{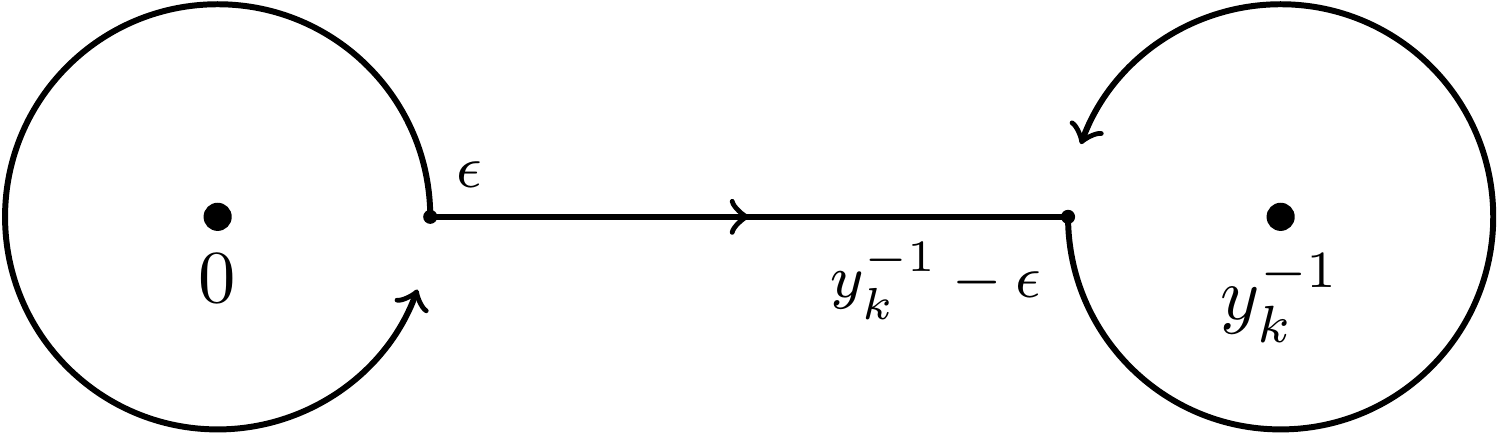}
\end{center}
\caption{\label{fig.cycI}The support $\eta_k$ of the locally-finite cycle $\check{\gamma}_k$ on the left-hand side and its regularised version on the right-hand side.}
\end{figure}

In the univariate case, the $r+1$ linear factors of the twist take the form $\{x,L_i(x)= (1-y_j x)\}$. They define a set of points $\Sigma = \{ 0,y_j^{-1}|0\leq j\leq r\}$ that we assume to be ordered $y_{i}^{-1}<y_{j}^{-1}$ for $i<j$ (for simplicity, we assume that these points are real). Then the basis of locally-finite dual cycles can be chosen to be supported on $r$ open intervals between the points $y_i^{-1}$.  We consider specifically the two bases
\begin{align}
\label{defeta}
\check{\gamma}_j &=\eta_j\otimes \Phi^{-1}_{\eta_j} \text{ with } \eta_j = (0,y_j^{-1})]\,, \\
\check{\tilde{\gamma}}_j &=\tilde{\eta}_j\otimes \Phi^{-1}_{\tilde{\eta}_j} \text{ with }{\tilde{\eta}}_j =(y_{j-1}^{-1},y_{j}^{-1})\,,\label{interval2}
 \end{align}
with  the orientations as indicated in the left panel of figure \ref{fig.cycI}.  Note, that these intervals are deformed such that they do not intersect the other punctures $y_i^{-1}$ that are taken out of $X$. We fix the branch of the twist on these intervals by the prescription
\begin{align}\label{presarg}
\text{arg}[L_j(x)]=\begin{cases} 0 & \text{ if } 1\leq j \leq k \\ -\pi & \text{ if } k+1\leq j\leq m \end{cases}\text{ on the interval $(y_k^{-1},y_{k+1}^{-1})$}\,,
\end{align}
This choice of arguments is equivalent to saying that we define the twist on the lower half-plane, i.e., we analytically continue around the points $y_j^{-1}$ through the lower half-plane, picking up a factor $e^{-i\pi a_j}$ along the way. To obtain the compactly supported twisted cycles with support $\eta_{j,c}$ we can \textit{regularise} the 
open intervals $\eta_j$ by taking $\epsilon$-circles around the singularities as depicted in the right panel of figure \ref{fig.cycI}. Specifically, we consider twisted cycles 
\begin{align}
\label{regcyc}
\gamma_j= \eta_{j,c}\otimes \Phi|_{\eta_{j,c}} =\frac{S_{\epsilon}(0)}{d_0} \otimes \Phi|_{S_{\epsilon}(0)}+ (\epsilon, y_{j}^{-1}-\epsilon)\otimes \Phi|_{(0+\epsilon, y_{j}^{-1}-\epsilon)}- \frac{S_\epsilon(y_{j}^{-1})}{d_{j}}\otimes \Phi|_{S_\epsilon(y_{j}^{-1})}\, .
\end{align}
The circles $S_\epsilon$ are oriented positively (anti-clockwise) and the branches are chosen by analytically continuation along these circles, taking into account the assignments of branches for the interval $ [\epsilon, y_{j}^{-1}-\epsilon]$ in eq.~(\ref{presarg}). The normalising factors $d_j$ guarantee, that the boundary of the regularised twisted cycle vanishes:
\beq\bsp
    \partial \left(\eta_{j,c} \otimes \Phi|_{\eta_{j,c}} \right) &\,= \frac{\epsilon}{d_0} \left(c_{0}-1\right)+ \left(y_j^{-1}-\epsilon -\epsilon\right) - \frac{y_j^{-1}-\epsilon}{d_j} \left(c_j-1\right)
 =0\, . 
\esp\eeq
The intervals $\tilde{\eta}_j$ of eq.~(\ref{interval2}) can be regularised in a similar manner. 

\subsection{Intersection numbers of twisted cycles} 
\label{sec.hom.2}

Consider a twisted cycle $|\gamma_B]\in H_n(X, \check{\mathcal{L}}_{\underline{a}} ) $ and a dual twisted cycle $[\check{\gamma}_A|\in H_n^{\text{lf}} (X, {\mathcal{L}}_{\underline{a}})$, both  decomposed in embedded cycles as in eqs.~(\ref{twistedcycel}) and (\ref{duatwistedcycle}). Their intersection number is 
\begin{align}
\label{22.6}
[\check{\gamma}_A| \gamma_B ] =\sum_{v}[\check{\gamma}_A| \gamma_B ]_v =\sum_{\triangle, \square, \{v\} \in \triangle \cap \square} c_{\triangle} \,d_{\square}\, \Phi_\triangle|_v\,\Phi_{\square}^{-1}|_v\,I_v(\triangle, \square)\, 
\end{align}
with  the sum  taken over all points $v$, where the simplices $\triangle$ and $\square$ intersect. $I_v(\triangle, \square)$ is the topological intersection number at $v$: 
\begin{align}
    I_v(\triangle,\square)= \begin{cases} +1& \triangle \text{ and } \square \text{ intersect with positive orientation at } v:   \adjincludegraphics[scale=0.2,valign=c]{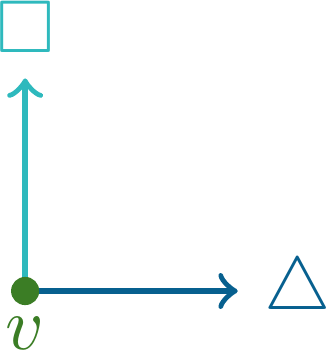} \\
    -1& \triangle \text{ and } \square \text{ intersect with negative orientation } v:   \adjincludegraphics[scale=0.2,valign=c]{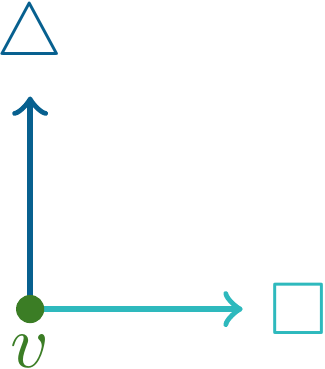} \end{cases}\, . 
\end{align} 
We call $[\check{\gamma}_A| \gamma_B ]_v$ the \textit{local intersection number at $v$}
We now briefly review how to apply the formula in eq.~(\ref{22.6}) to evaluate these intersection numbers between twisted cycles. We discuss separately the univariate and multivariate cases.

\subsubsection{Univariate intersection numbers}
There are three distinct configurations where  the twisted cycle and the dual locally-finite cycle meet. We review each case in turn following ref.~\cite{kita_intersection_1994-2}.

\paragraph{Local intersection near a point $y_i^{-1}$ with agreeing branches of the twist.}
\begin{figure}[!t]
\begin{center}
\includegraphics[scale=0.35]{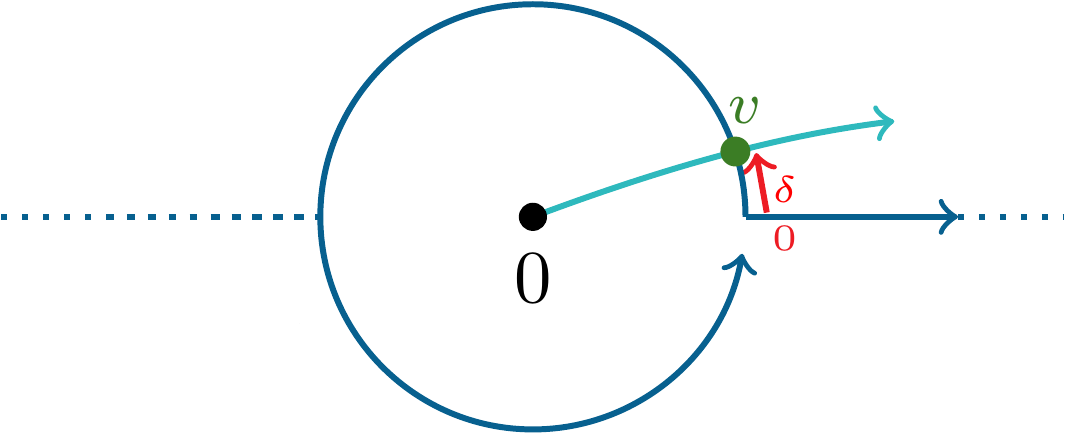}\hskip 1cm
\includegraphics[scale=0.35]{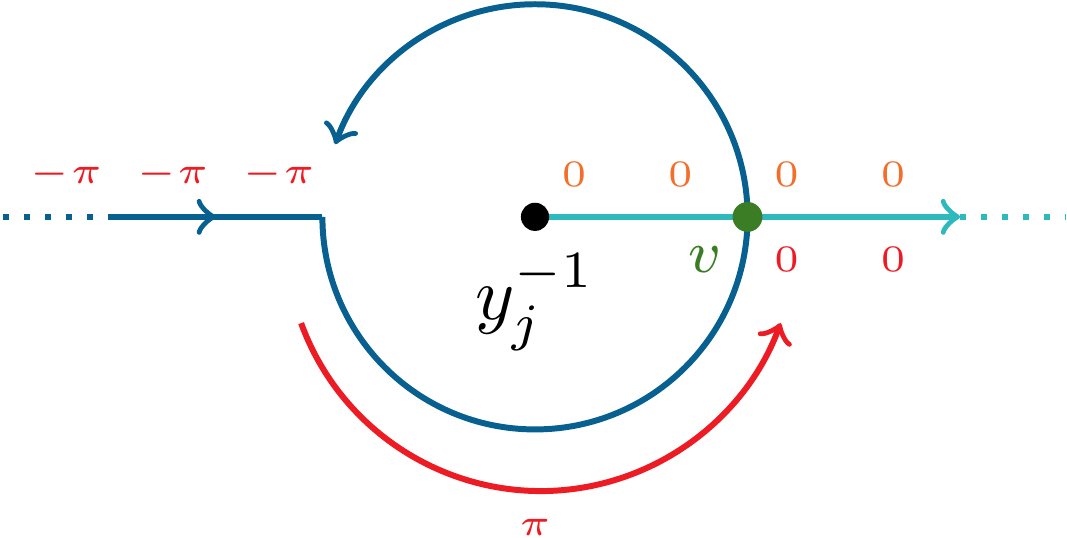}
\end{center}
\newpage
\caption{Two examples for a local intersection at $v$ near a puncture  $y_i^{-1}$ where the branches chosen for the twist agree on the twisted cycle and the dual twisted cycle.}
\label{fig.51}
\end{figure}

For a point $v$ near the puncture $y_j^{-1}$, where the branches loaded on the cycle and the dual cycle agree, the local intersection number is just $\pm \frac{1}{d_j}$. A simple case with a contribution from such a local intersection number is depicted in the left panel of figure \ref{fig.51}: One of the points where the interval $\eta_j=(0, y_{j}^{-1})$ meets its regularised version $\eta_{j,c}$ is at  $v$ near $0$, where the twist (specifically its factor $L_0^{a_0}$(x)) is analytically continued  along the same path  for both the twisted cycle $\gamma_j$ supported on $\eta_{j,c}$ and its dual $\check{\gamma}_j$ supported on $\eta_j$. We obtain:
\begin{align}
    [\check{\gamma}_j|\gamma_j]_v&= \frac{1}{d_0} e^{i\delta a_0}e^{-i\delta a_0} (-1)= -\frac{1}{d_0}\, . 
\end{align}
In the right panel of figure~\ref{fig.51}, the only local intersection between the locally finite twisted cycle $\check{\tilde{\gamma}}_{j+1}$ supported on $\tilde{\eta}_{j+1}=(y_j^{-1},y_{j+1}^{-1})$ and the twisted cycle $\tilde{\gamma}_j$ supported on the regularised interval $\tilde{\eta}_{j,c}$ is depicted. Besides $L_{j}$, all factors of the twist agree on the intervals $(y_j^{-1},y_{j+1}^{-1})$ and $(y_{j-1}^{-1},y_{j}^{-1})$. According to the prescription in eq.~\eqref{presarg}, the factor $L_{j}$ has the following arguments on these intervals: 
\begin{align}\label{presarg2}
\text{arg}[L_{j}(x)]=\begin{cases} 0 & \text{ on the interval $(y_j^{-1},y_{j+1}^{-1})$}\,,
\\ -\pi & \text{ on the interval $(y_{j-1}^{-1},y_{j}^{-1})$}\,.
\end{cases}
\end{align}
Due to the analytical continuation of the twist along $S_{\epsilon}(x_{j+1})$, the argument of its factor $L_{j}$ receives an extra $\pi$ at $v$ on $\tilde{\eta_{j,c}}$. Thus, the branches of the twist assigned to ${\tilde{\eta}}_{j,c}$ and ${\tilde{\eta}}_{j+1}$ agree at $v$ and we obtain:
\begin{align}
\label{23.8}
[\check{\tilde{\gamma}}_{j+1}|\tilde{\gamma}_{j}]_v=  \frac{1}{d_{j}}\, . 
\end{align}
\paragraph{Local intersection near a point $y_i^{-1}$ with disagreeing branches of the twist.}

\begin{figure}[!t]
\begin{center}
\begin{center}
\includegraphics[scale=0.35]{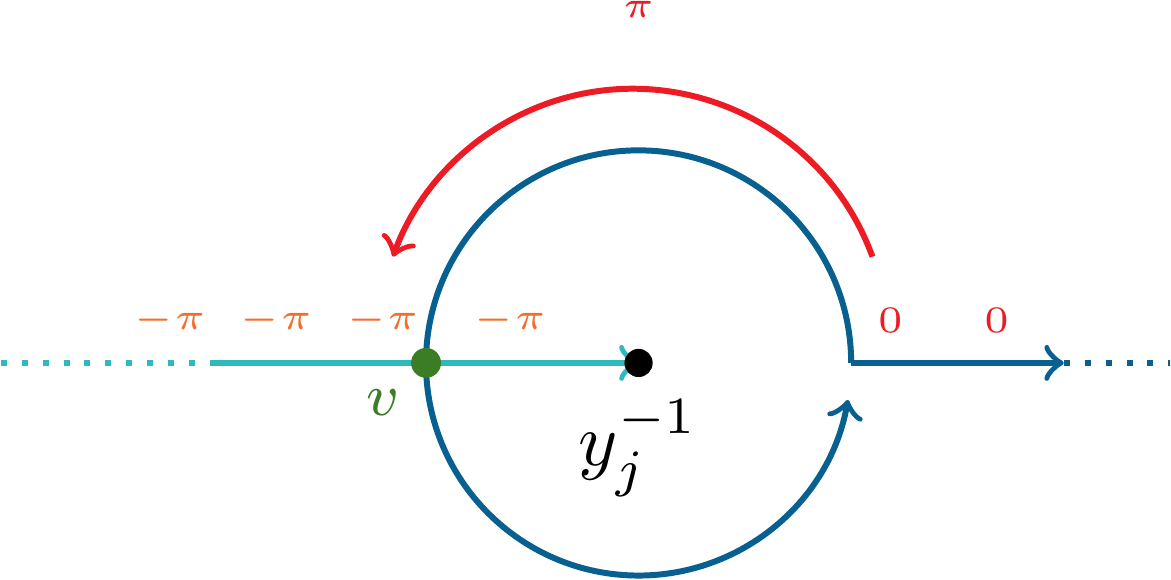}
\end{center}
\end{center}
\caption{\label{fig.6}An examples for a local intersection at $v$ near a puncture  $y_i^{-1}$ where the branches chosen for the twist do not agree on the twisted cycle and the dual twisted cycle. }
\end{figure}
An example for a local intersection number, where we pick up an additional factor $c_j$ due to differing branch choices, is the local intersection number between $\check{\tilde{\gamma}}_{j-1}$ and $\tilde{\gamma}_j$ at the point $v$ near $y_j^{-1}$ (see figure~\ref{fig.6}):  The arguments of $L_j$ on the respective intervals are given in eq.~(\ref{presarg2}). The  analytic continuation of the twist along $S_\epsilon(y_j^{-1})$ through the upper half-plane gives an extra argument of $\pi$ at $v$ on ${\tilde{\eta}}_{j,c}$ for the factor $L_j$ of $\Phi$. Thus
\begin{align}
\label{23.7}\Phi|_{v\in{\tilde{\eta}}_{j,c}}\Phi^{-1}|_{v\in \tilde{\eta}_{j-1}}= e^{2\pi i a_j}=c_j\,,
\end{align}
and 
\begin{align}
\label{extwist1}
[\check{\tilde{\gamma}}_{j-1}|\tilde{\gamma}_j]_v&=  \frac{c_j}{d_j}.
\end{align}
The latter two examples illustrate why the choice of arguments as indicated in eq.~\eqref{presarg} is equivalent to saying that the functions are defined on the lower half-plane: As long as we only analytically continue in the lower half-plane, we follow the analytical continuation indicated by this choice of arguments and stay on the same branch. If we analytically continue through the upper half-plane, we land on different branches.

\paragraph{Local intersection on the interval between two points}

\begin{figure}[!t]
\begin{center}
\includegraphics[scale=0.35]{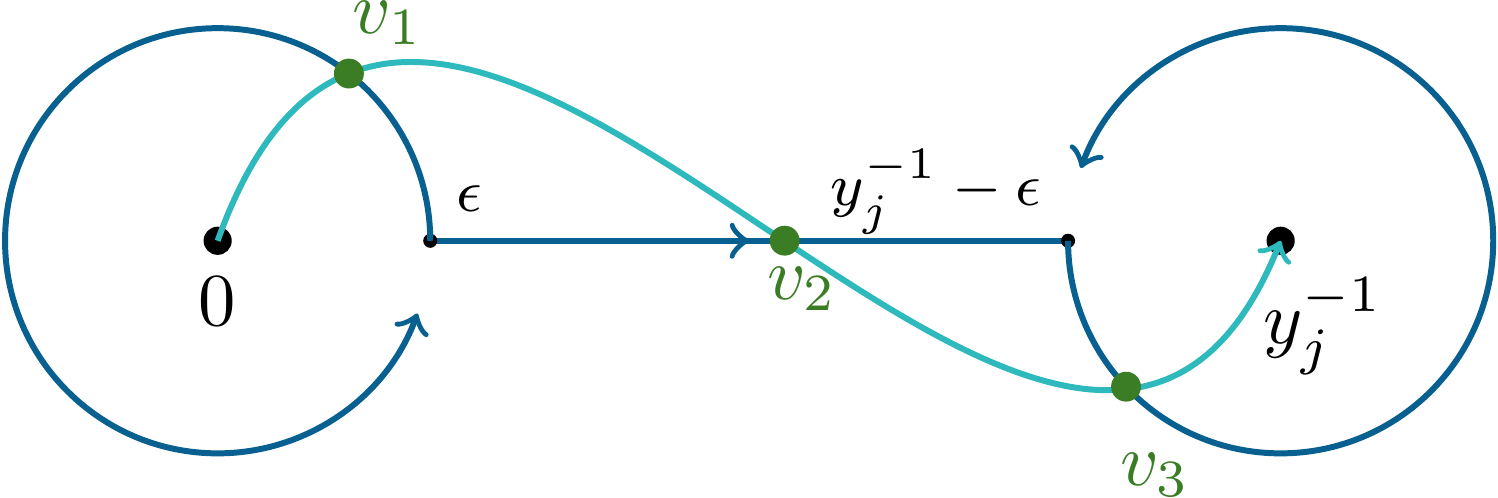}
\end{center}
\caption{The self-intersection between the twisted regularised cycle \textcolor{UniBlau}{$\gamma_j$} and the dual cycle \textcolor{Aquamarine}{$\check{\gamma}_j$}.}
\label{fig.4}
\end{figure}

The local intersection number at a point on the interval between two punctures $y_i^{-1}$ and $y_{i+1}^{-1}$ is just  the topological intersection number $\pm 1$. This case contributes for example to the self-intersection number of a twisted cycle $\gamma_j$, as depicted in figure \ref{fig.4}. Here, the locally finite dual cycle can be deformed such that it meets the regularised cycle in the three points $v_1,v_2$ and $v_3$ and the intersection number is then:
\begin{align}
\label{diagoint}
[\check{\gamma}_j|\gamma_j]_{v_1}+[\check{\gamma}_j|\gamma_j]_{v_2}+[\check{\gamma}_j|\gamma_j]_{v_3}= -\frac{1}{d_j}-1-\frac{1}{d_{j+1}} =-\frac{d_{j,j+1}}{d_jd_{j+1}} \, . 
\end{align}

\subsubsection{Multi-variate case} 

In refs.~\cite{aomoto_theory_2011,kita_intersection_2006-2}  methods  for computing the intersection numbers between twisted cycles supported on higher-dimensional chambers were developed. The basic idea is to deconstruct the chambers into one-dimensional building blocks whose intersection numbers can be  computed as discussed in the first part of this section. 

As in the one-dimensional case, the first step is to determine a basis of chambers for the locally-finite cycles and make a choice for the arguments of $L_j$ in each chamber. In the following diagrams, we indicate with ``$+$'' that we choose $\text{arg}\left[L_j\right]= 2n\pi$ in a given region and with ``$-$'' that we choose an uneven multiple of $\pi$ for $\text{arg}\left[L_j\right]$ in that region. For now, let us assume that we want to compute the intersection number between two specific cycles $\check{\gamma}_E$ and $\gamma_D$ that are supported on neighbouring chambers $E,D$ and make a choice for the arguments that is beneficial for this specific computation. Let $H_{i_1},\dots H_{i_q}$ be the hyperplanes meeting normally along the intersection of $E$ and $D$ and $H_P=H_{i_1}\cap \dots \cap H_{i_p}$ its interior. In this setup we assign the branches
\begin{align}
\label{argL}
\text{arg}[ L_{i_1}(\underline{x})]=\dots =\text{arg} [L_{i_q}(\underline{x})]=0&\text{ on } D\,,\\
\text{arg} [ L_{i_1}(\underline{x})]=\dots =\text{arg} [ L_{i_p}(\underline{x})]=-\pi \text{ and } \text{arg} [ L_{i_{p+1}}(\underline{x})]=\dots =\text{arg} [ L_{i_q}(\underline{x})]=0 &\text{  on } E\, . 
\end{align}
We denote the intersection number for this specific choice of branches by $\llbracket\eta|\gamma\rrbracket$ and it can be computed by \cite{yoshida_hypergeometric_1997} 
\begin{align}
\label{intnumgen}
\llbracket\check{\gamma}_E|\gamma_D\rrbracket=(-1)^{n-p} \frac{c_{i_1}\dots c_{i_p}}{d_{i_1}\dots d_{i_p}} \left(1+ \sum_{k=1}^{n-p}\sum_{{i_{p+1}}\leq j_1\leq \dots \leq j_k\leq {i_q}} \frac{1}{d_{j_1}\dots d_{j_k}}\right)\, . 
\end{align}
 The summation over $j_i$ is taken such that we only consider $j_1\leq \dots \leq j_k$ with $D\cap E \cap\bigcap_{j_i\in j_1\dots j_k} H_{j_i}\neq 0$. For $n=2$, the self-intersection number can also be read off from a pictorial representation of a cycle by assigning to the barycenter a $1$, to each edge a factor $\frac{1}{d_i}$ and to each vertex between two lines a factor of $\frac{1}{d_id_j}$ and summing over all these factors. The case where more than two lines meet is degenerate and discussed below. The intersection number $[.|.]$ of an arrangement with any  choice of arguments is then related to the intersection number $\llbracket.|.\rrbracket$ by a prefactor of powers of $c_i$ (due to analytic continuations similar to the ones discussed for the univariate case). Note that for the self-intersection number, this prefactor is always equal to unity, since we change the branch in the same way for the cycle and its dual. Thus, it is convenient to work in a homology basis, where the cycles do not intersect each other and the intersection matrix is diagonal.

\begin{figure}[!t]
\begin{center}
\includegraphics[scale=0.25]{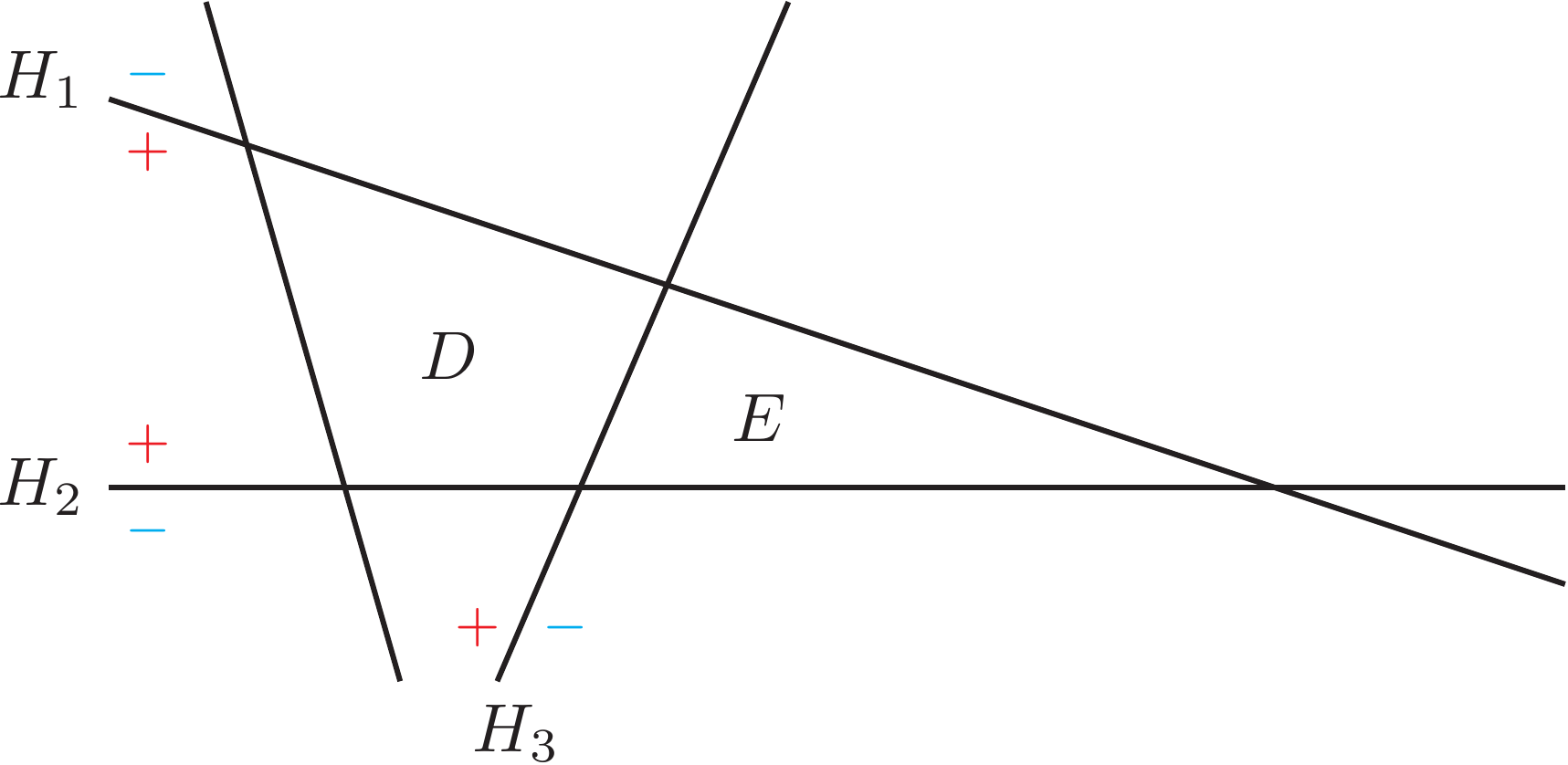}\, \, \, \, \, \,
\includegraphics[scale=0.25]{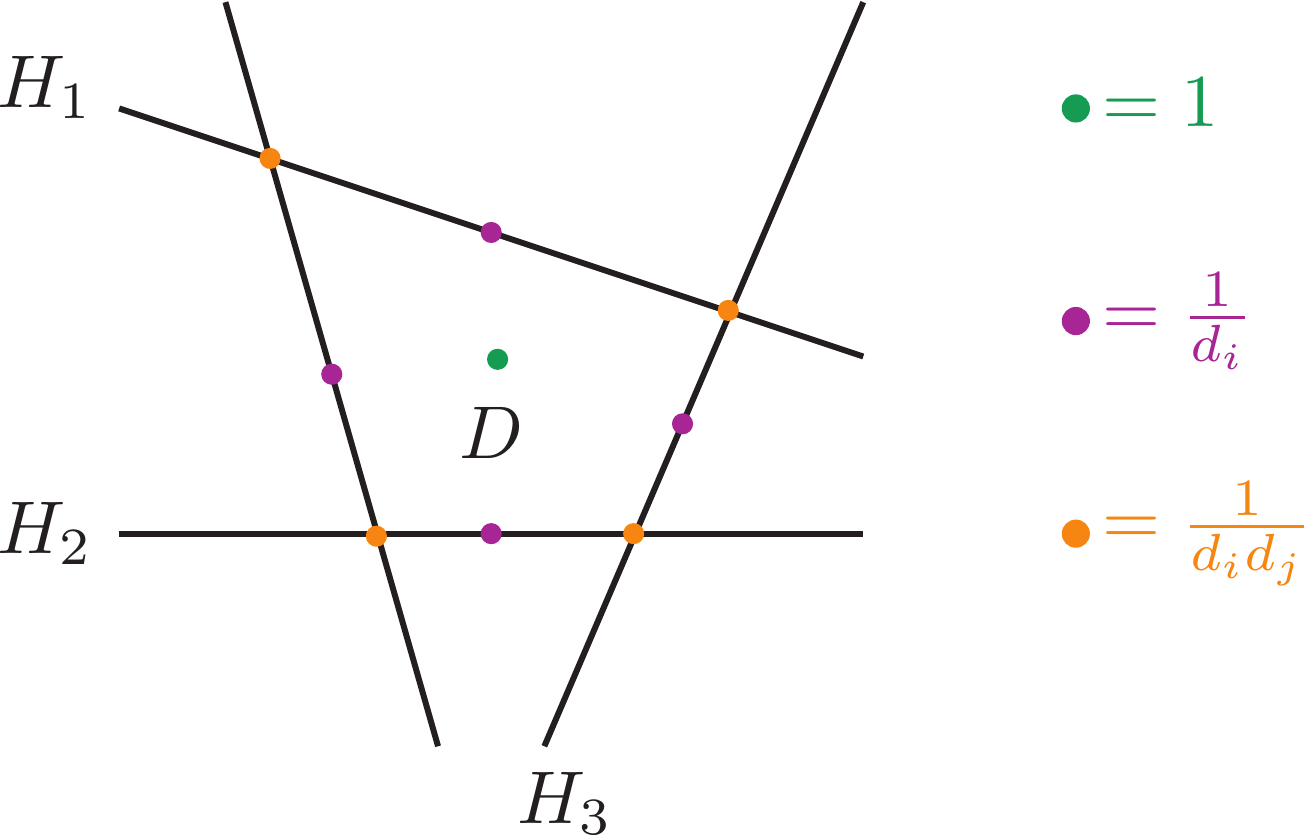}
\end{center}
\caption{\label{fig.22}An arrangement of hyperplanes bounding the two chambers $D,E$ is depicted on the left panel. The right panel illustrates the computation of a self-intersection number in $n=2$.}
\end{figure}

\begin{figure}[!t]
\begin{align*}
\adjincludegraphics[scale=0.12,valign=c]{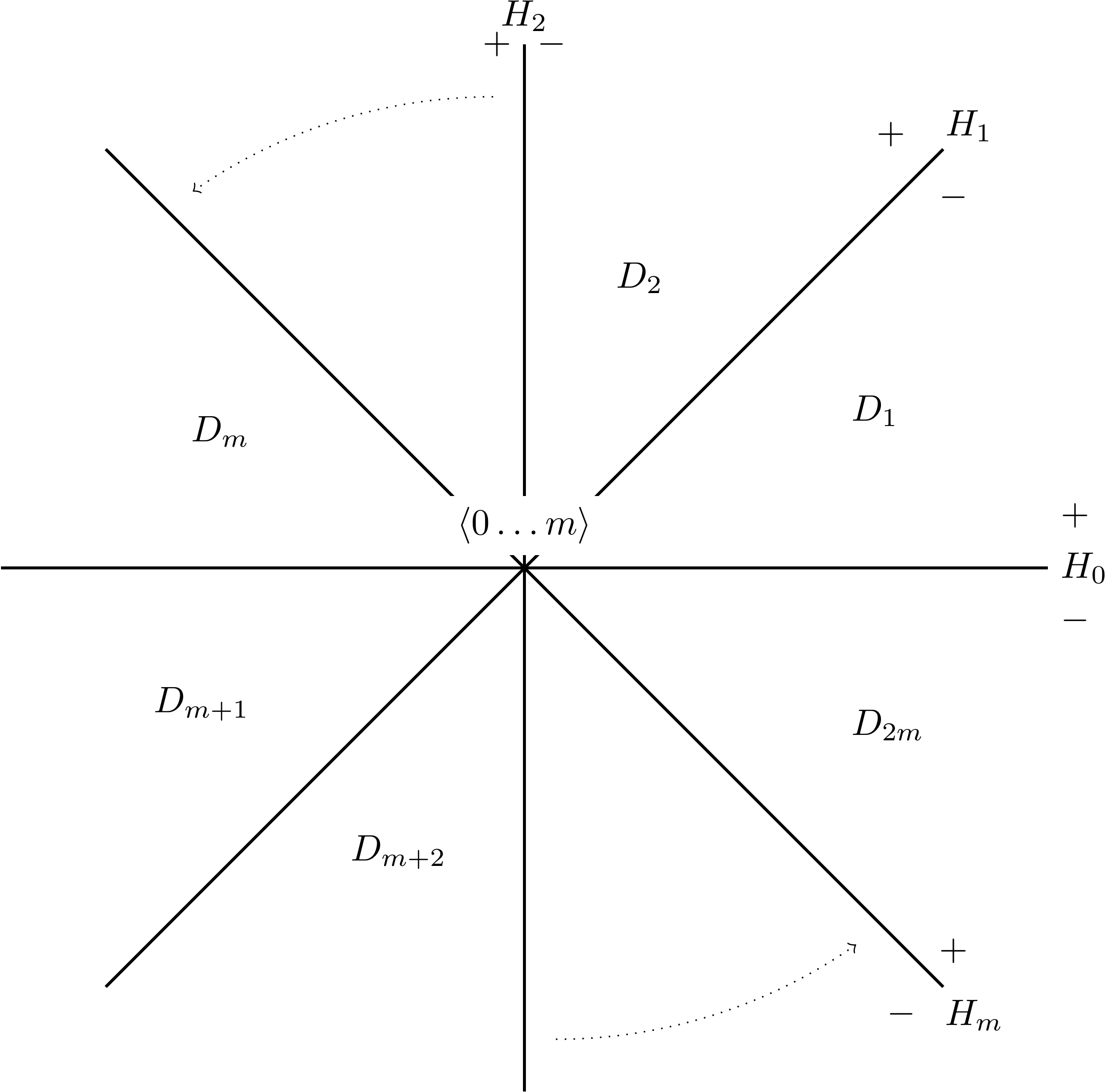}\hspace{5em}
\adjincludegraphics[scale=0.2,valign=c]{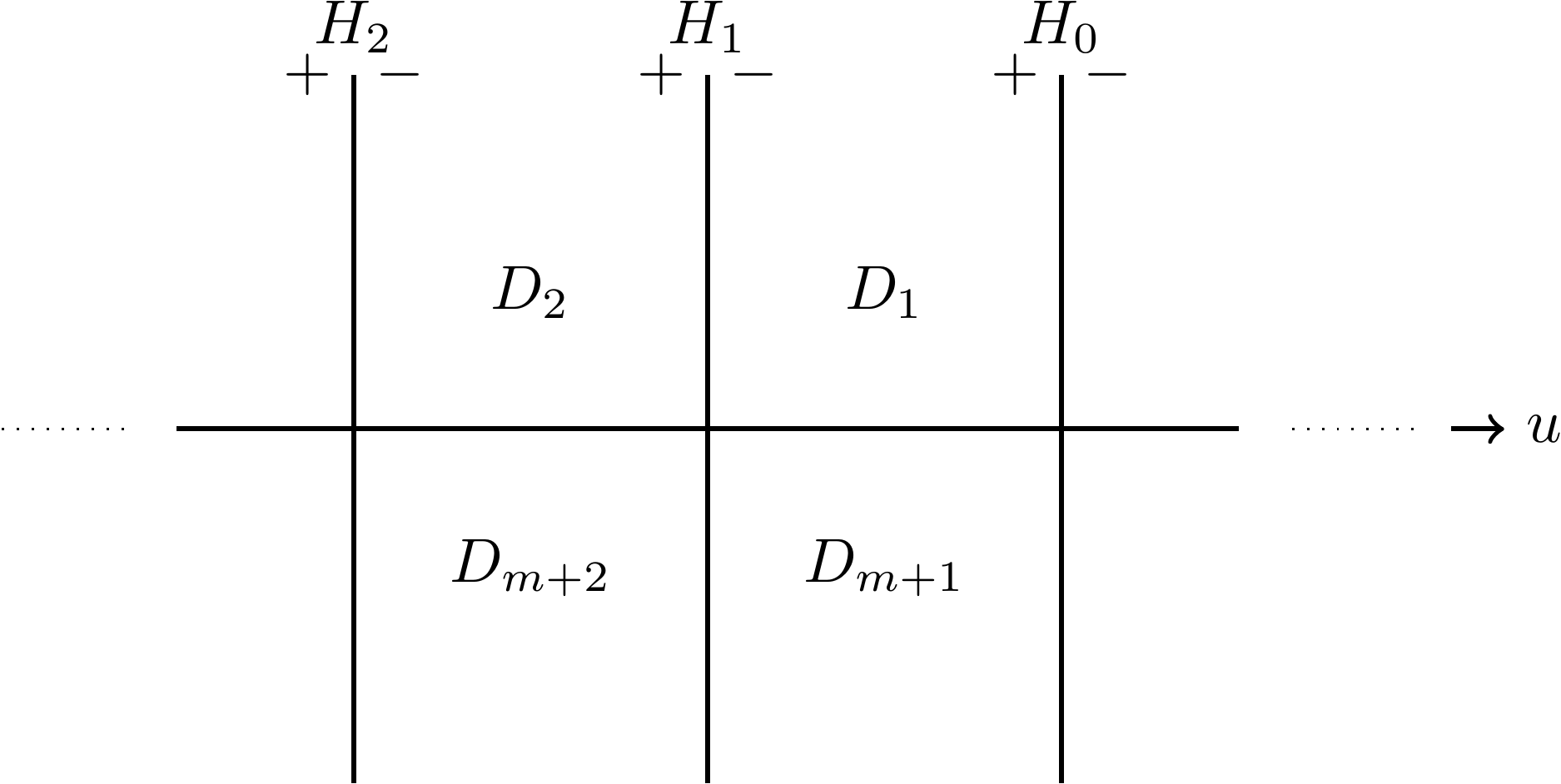}
\end{align*}
\caption{\label{fig.lines}A degnerate arrangement of $m$ lines depicted in a projection into $\mathbb{R}^2$ on the left-hand side and their blow-up on the right-hand side. }
\end{figure}

 \paragraph{Example:} An example for such an arrangement is given in the left panel of figure \ref{fig.22}. Here, $H_P=H_3$ is the interior of the intersection of $D,E$ and it is intersected by the hyperplanes $H_1,H_2$ to which the following arguments of the factors $L_i$ are assigned
\begin{align}
\text{arg}[L_1(\underline{x})]=\text{arg}[L_2(\underline{x})]=\text{arg}[L_3(\underline{x})]=0\text{ on } D\,,\\
\text{arg}[L_1(\underline{x})]=\text{arg}[L_2(\underline{x})]=0 \text{ and } \text{arg}[L_3(\underline{x})]=-\pi\text{ on } E\,.
\end{align}
Then, by (\ref{intnumgen}): 
\begin{align}
[\check{\gamma}_E|\gamma_D]= \llbracket\check{\gamma}_E|\gamma_D\rrbracket= -\frac{c_3}{d_3} \left(1+\frac{1}{d_1}+\frac{1}{d_2}\right) = -\frac{c_3d_{12}}{d_1d_2d_3}\, . 
\end{align}
Similarly, one obtains the self intersection number of $\gamma_1$:
\begin{align}
    [\check{\gamma}_D|\gamma_D]&= 1+ \frac{1}{d_1}+\frac{1}{d_2}+\frac{1}{d_3}+\frac{1}{d_4}+\frac{1}{d_1d_2}+\frac{1}{d_2d_4}+\frac{1}{d_4d_3}+\frac{1}{d_3d_4} =\frac{d_{13}d_{24}}{d_1d_2d_3d_4}\, . 
\end{align}
As illustrated in the right panel of figure \ref{fig.22} one can also obtain this result graphically by adding up contributions from different kinds of boundaries. 
\paragraph{Degenerate arrangements and relative homology.} 
The methods above are sufficient as long as the hyperplanes are in general position. Strategies for the treatment of degenerate arrangements (e.g., including vertices where more than two hyperplanes meet) are given in ref.~\cite{yoshida_intersection_2000-1}. 
The general idea is to blow-up the degenerate vertex into an arrangement in general position. Since the construction is more involved, we will be brief and focus only on the case of $m$ lines meeting depicted in a projection of $\mathbb{C}^2$ into $\mathbb{R}^2$, on the right panel of figure~\ref{fig.lines}. We refer to ref.~\cite{yoshida_intersection_2000-1} for the treatment in a more general situation.
We blow-up the vertex $\langle 1\dots m\rangle$ to be the hyperplane $H_u$ along which a new coordinate $u$ is introduced as depicted in the left panel of figure~\ref{fig.lines}. The hyperplane $H_u$ is loaded with $\sum_{i=0}^m a_i$, i.e. $c_u= c_{0\dots m}$. Then we can compute intersection numbers of cycles supported on the chambers depicted on the right-hand side as sums of their local intersection numbers along the hyperplane $H_u$. 

As an example we consider the local intersection number contributing to the self-intersection of the twisted cycle supported on $D_1$ (of the left panel of figure \ref{fig.lines}). It has a contributions from the barycenter $B_1$ and from the two vertices $\langle u,0\rangle$ and $\langle u,1\rangle$ which sum up to 
\begin{align}
\label{selfintdeg}
    [\check{\gamma}_1|\gamma_1]_{\langle 0,\dots, m\rangle} = \frac{1}{d_{u}}+\frac{1}{d_ud_0}+\frac{1}{d_ud_1}=\frac{d_{01}}{d_0d_1}\frac{1}{d_u}=\frac{d_{01}}{d_0d_1} \frac{1}{d_{0\dots m}} \, . 
\end{align}

In cases where we consider hyperplanes related to poles as well as to multi-valued points, a treatment with relative homology is required. For the case of Lauricella-functions (i.e. only the univariate non-degenerate case) a method to find a suitable basis of relative twisted cycles  and compute intersection numbers is given in ref.~\cite{matsumoto2019relative}.

\section{Twisted periods and intersection matrices of specific hypergeometric functions}
\label{examplesCoHom}
In this appendix we present the explicit results for the matrices $\bP_{\underline{a}}$, $\bC_{\underline{a}}$ and $\bH_{\underline{a}}$ for the Lauricella $F_D^{(r)}$ and generalised hypergeometric ${}_{p+1}F_p$ functions. These matrices are sufficient to computed the single-valued period matrix $\bP_{\underline{a}}^{\text{sv}}$ for all one-loop integrals and all one-parameter ladder integrals. 

\subsection{The Lauricella $F_D^{(r)}$ function}
\label{app.perlaur}

Throughout this section, we use the notations 
\beq\bsp
\underline{a} &\,= (a,b_1,\ldots,b_r,c)\,, \\
\underline{y} &\,=(y_1,\ldots,y_r)\,,\\
\underline{y}^{(k)} &\,=\left(\frac{y_1}{y_k},\ldots,\frac{y_{k-1}}{y_k},\frac{1}{y_k},\frac{y_{k+1}}{y_k},\ldots,\frac{y_r}{y_k}\right)\,.
\esp\eeq

\paragraph{The period matrix $\bP_{\underline{a}}$.}
From eq.~\eqref{eq:Lauricella1}, we get:
 \begin{align}
 \big(\bP_{\!\underline{a}}\big)_{11} = \langle \varphi_{r,1} |\gamma_{r,1}] = \cF_D^{(r)}(\underline{a};\underline{y})\, . 
\end{align}
Integrating $\langle \varphi_{r,1}|$ over another cycle $\gamma_{r,k}$ we obtain the period
\begin{align}
   \big(\bP_{\!\underline{a}}\big)_{1k} =  \langle \varphi_{r,1} |\gamma_{r,k}]= y_{k-1}^{-a} \cF_D^{(r)}\left(a, b_1, \dots,1+ a-c, \dots,b_r,  1-b_{k-1}+a;  \underline{y}^{(k-1)}\right)\,,
\end{align}
whereas integrating another co-cycle $\varphi_{r,k}$ over $\gamma_{r,1}$ we obtain
\begin{align}
    \big(\bP_{\!\underline{a}}\big)_{k1} =   \langle \varphi_{r,k} |\gamma_{r,1}]= \cF_D^{(r)}\left(a,b_1,\dots, b_{k-2}, b_{k-1}+1, b_{k} ,\dots, b_{r},c+1;\underline{y} \right)\, . 
\end{align}
In the most general case we obtain:
\begin{align}
\label{laurper}
  &      y_{k-1}^{a}\, \big(\bP_{\!\underline{a}}\big)_{lk} \,=       y_{k-1}^{a} \,  \langle \varphi_{r,l} |\gamma_{r,k}]\\
\nonumber   & \,=\begin{cases}
 \cF_D^{(r)}\left(a, b_1,\dots, b_{l-1}+1,\dots, b_{k-2}, a-c, b_{k}, \dots, b_{r} ,1-b_{k-1}+a;\underline{y}^{(k-1)}\right)\,,&l\neq k\,, \\
  \cF_D^{(r)}\left(a, b_1,\dots, b_{k-2}, a-c, b_{k}, \dots, b_{r} ,-b_{k-1}+a;\underline{y}^{(k-1)}\right)\,,&l= k \,.
\end{cases} 
\end{align}

\paragraph{The cohomology intersection matrix $\bC_{\underline{a}}$.}
The intersection matrix between the twisted co-cycles in eq.~\eqref{bascocyLau} is 
\begin{align}
\bC_{r}&=\left(
\begin{smallmatrix}
   \frac{1}{a}+\frac{1}{c-a} &\dots & \frac{1}{a} &
   \frac{1}{a} \\
 \frac{1}{a} &
   \frac{1}{a}-\frac{1}{b_1} & \dots &\frac{1}{a} \\
   \vdots &\ddots &\ddots &\vdots \\
 \frac{1}{a} &\dots&\dots&
   \frac{1}{a}-\frac{1}{b_{r}} \\
\end{smallmatrix}
\right)\, . 
\end{align}

\begin{figure}[!t]
\begin{center}
\includegraphics[scale=0.3]{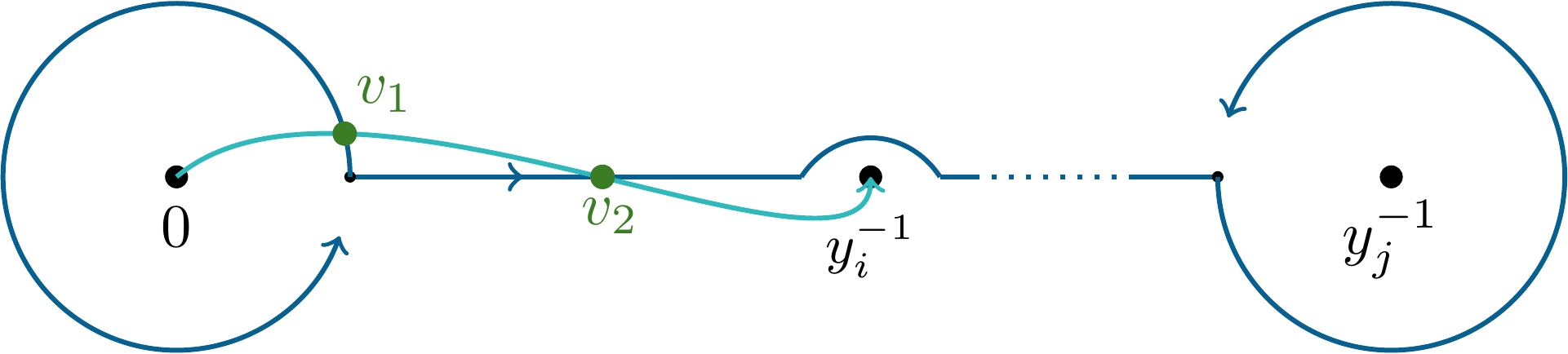}
\end{center}
\caption{\label{fig.19a}The two intersections of   $[\check{\gamma}_i|\gamma_j]$.}
\end{figure}

\paragraph{The homology intersection matrix $\bH_{\underline{a}}$.}
The intersection matrix for the homology group $H_1(\mathbb{C}-\Sigma_r, \cL_{\underline{a}})$ associated to a Lauricella hypergeometric function $\mathcal{F}_D^{(r)}$ with the basis of twisted cycles as in eq.~\eqref{laurcyc.1} is given by
\beq\bsp
\label{intGenLa}
  \bH_{\underline{a}}=&\,  \left(
\begin{smallmatrix}
 -\frac{d_{01}}{d_0d_1}&-\frac{c_0}{d_0}& \dots &
  -\frac{c_0}{d_0} \\
-\frac{1}{d_0} & -\frac{d_{02}}{d_0d_2}&\dots  & \vdots \\
 \vdots& \ddots& \ddots &\vdots\\-\frac{1}{d_0}  & \dots & \dots &-\frac{d_{0r}}{d_rd_0}\\
\end{smallmatrix}
\right)
=  \left(
\begin{smallmatrix}
 \frac{i }{2} (\frak{c}(
   a)+\frak{c}( c-a))&
    \frac{i}{2}  (\frak{c}(
   a)+i) & \dots &  \frac{i}{2}  (\frak{c}(
   a)+i)   \\
 \frac{1}{2} ( 1+i \frak{c}(
   a))& \frac{i}{2}  (\frak{c}
   (  a)-\frak{c}(
   b_1))&\dots  & \vdots \\
 \vdots& \ddots& \ddots &\vdots\\
 \frac{1}{2} ( 1+i \frak{c}(
   a)) & \dots & \dots &\frac{i}{2} 
   (\frak{c}( a)-\frak{c}(
   b_r))\\
\end{smallmatrix}
\right)\, ,
\esp\eeq
where we introduced the shorthand $\frak{c}(x)=\cot(\pi x)$.
The intersection number on the diagonal are computed as in eq.~\eqref{diagoint}. The upper triangular entries are  sums of two local intersection numbers
\begin{align}
    [\check{\gamma}_{r,i}|\gamma_{r,j}]=-\frac{1}{d_0}-1 =-\frac{c_0}{d_0} \text{ for } i<j
\end{align}
at intersections $v_1$ and $v_2$ as indicated in figure \ref{fig.19a}.

We have verified numerically that the Riemann bilinear relations from eq.~\eqref{EQU.24} hold for the period and intersection matrices given in this section for $r=0,1,2$. Note that to evaluate the period matrix for the Appell $F_1$ function ($r=2$) numerically, analytical continuations as given in ref.~\cite{bezrodnykh_analytic_2017} are necessary.
We can use these matrices to express the single-valued analogues of Lauricella's $F_D^{(r)}$ functions as bilinears in $F_D^{(r)}$ function.
For $r=1$, we reproduce the results for the single-valued analogue of Gauss' ${}_2F_1$ function of ref.~\cite{brown_lauricella_2019}. Note that ref.~\cite{brown_lauricella_2019} works with a different basis of cycles where the homology intersection matrix~$  \bH_{\underline{a}}$ is diagonal.

\subsection{Generalised hypergeometric ${}_{p+1}F_p$ functions}
\label{perfp}
Throughout this section we use the notations 
\beq\bsp
\underline{a} &\,= (a_0,\ldots,a_p)\,,\\
\underline{b} &\,= (b_1,\ldots,b_p)\,,\\
\underline{\mu} &\,= (\underline{a},\underline{b})\,,\\
\underline{a}_k &\,=  \underline{a}_k=(1+a_k-b_k,1+a_1-b_k,\dots, 1+a_p-b_k, 1+a_0-b_k)\,,\\
\underline{b}_k &\,= \underline{b}_k=(1-b_k+b_1, \dots, 1-b_{k}+b_{k-1},1-b_{k}+b_{k+1}, \dots, 1-b_{k}+b_p, 2-b_{k})
\,,\\
 \underline{{a}}_{kl}&\, =(1+a_l-b_l-\delta_{kl},1+a_1-b_l-\delta_{kl},\dots, 1+a_p-b_l-\delta_{kl}, 1+a_0-b_l-\delta_{kl})\,,\\
    \underline{{b}}_{kl}&\, =(1-b_l+b_1-\delta_{lk}, \dots, 1-b_{l}+b_{l-1}-\delta_{lk},1-b_{l}+b_{l+1}-\delta_{lk},\\
   & \, \, \,  \, \, \,  \, \, \,  \, \, \, \dots, 2-b_l+b_k, \dots, 1-b_{l}+b_p-\delta_{lk}, 2-b_{l}-2\delta_{lk})\,.
\esp\eeq

\paragraph{The periods matrix $\bP_{\underline{\mu}}$.}
We give the period matrix for the generalised hypergeometric ${}_{p+1}F_p$ function from ref.~\cite{goto2014intersection}. The first line (i.e. integrations of the first basis co-cycle over different cycles) and the first column (i.e. integrations of the different basis co-cycles over the first basis cycle) of the intersection matrix are given by
\begin{align}
  \nonumber  \left(\bP_{\underline{\mu}}\right)_{11} =&\, {}_{p+1}\mathcal{F}_p(\underline{a},\underline{b}; y)\,,
    \\
   \left(\bP_{\underline{\mu}}\right)_{1k} =&\, (-1)^{k} e^{-i \pi (1+a_0+a_{k-1}-b_{k-1})}y^{1-b_{k-1}}\,{}_{p+1}\mathcal{F}_p(\underline{a}_{k-1},\underline{b}_{k-1}; y)\,,
 \nonumber\\
 \left(\bP_{\underline{\mu}}\right)_{k1} =&y {}_{p+1}\mathcal{F}_p (1+a_0,1+a_1,\dots, 1+a_p, 1+b_1, \dots, 2+b_{k-1},\dots, 1+b_p;y)\, . 
\end{align}
Most generally, we obtain:
\begin{align}
\left(\bP_{\underline{\mu}}\right)_{kl} =&(-1)^l y^{1-b_{l-1}-\delta_{lk}} e^{-i\pi(2+a_0+a_{l-1}-b_{l-1}-\delta_{kl})}{}_{p+1}\mathcal{F}_p (\underline{a}_{k-1,l-1},\underline{b}_{k-1,l-1};y)\,.
\end{align}

\paragraph{The cohomology intersection matrix $\bC_{\underline{\mu}}$.}
The intersection matrix is 
\begin{align}
\label{intF1A}
\bC_{\underline{\mu}}&=
\left(
\begin{smallmatrix} \prod_{i=1}^p \frac{b_i}{a_i (a_i-b_i)}& 0 & 0 &0 \\
 0 & - \frac{b_1}{a_0(a_0-b_1)} \prod_{i=1, i\neq 1}^p \frac{b_1-b_i}{(b_1-a_i)(a_i-b_i)} &0&0\\
 0&0&\ddots&0\\
 0&0&0& - \frac{b_p}{a_0(a_0-b_p)} \prod_{i=1, i\neq p}^p \frac{b_p-b_i}{(b_p-a_i)(a_i-b_i)}\\
\end{smallmatrix}
\right)\,.
\end{align}

\paragraph{The homology intersection matrix $\bH_{\underline{\mu}}$.}
The intersection theory for ${}_{p+1}F_p$ functions is discussed in refs.~\cite{mimachi_intersection_2010,goto2014intersection}. The case $p=1$ was already discussed in the context of Lauricella's $F_D^{(r)}$ function. We compute the homology intersection matrix for the ${}_3F_2$ function explicitly and deduce from that the generalisation to $p>2$. 

\underline{The case $p=2$:}
The projection of $X_{{}_3F_2}$ in $\mathbb{R}^2$ with a qualitative illustration of its parallel lines meeting the hyperplane at infinity is depicted in figure~\ref{fig.f32space}. The three chambers from eq.~\eqref{subf32} supporting the basis of cycles are denoted by $D_1,D_2,D_3$. 
Since these chambers do not meet, the intersection matrix is diagonal. Some of the local intersection numbers are degenerate (i.e., there are points where three lines intersect) and can be computed with eq.~\eqref{selfintdeg}. A similar configuration was considered in ref.~\cite{kita_intersection_2006-4}. We obtain for the  three non-zero elements of $\bH_{\underline{\mu}}$: 
\begin{align}
\label{h32}
    \left(\bH_{\underline{\mu}}\right)_{11}&=1+ \frac{1}{d_{1}}+\frac{1}{d_2}+\frac{1}{d_{4}}+\frac{1}{d_1d_2}+\frac{1}{d_2d_4}+\frac{d_{14}}{d_1d_4}\frac{1}{d_{014}}=\frac{d_{14}d_{0124}}{d_1d_2d_4d_{014}}\\
    \left(\bH_{\underline{\mu}}\right)_{22}&=1+\frac{1}{d_3}+\frac{1}{d_4} + \frac{1}{d_\infty} + \frac{1}{d_3d_4} + \frac{1}{d_4d_\infty}+\frac{d_{3\infty}}{d_3d_\infty}\frac{1}{d_{13\infty}}=
    \frac{c_3c_4d_{02} d_{0124}}{d_3 d_4d_{024} d_{01234}} \notag 
    \\
    \left(\bH_{\underline{\mu}}\right)_{33}&=1+\frac{1}{d_0}+\frac{1}{d_2} +\frac{1}{d_3} +\frac{1}{d_3d_0}+\frac{1}{d_2d_3} +\frac{d_{02}}{d_0d_2}\frac{1}{d_{02\infty}}=\frac{c_3d_{02}d_{14} }{d_0d_2d_3d_{134} }\, . \notag
\end{align}
Note that, since $\mu_\infty = -\sum_{i=0}^4\mu_i$, we have $c_\infty=c_{01234}^{-1}$.  The full intersection matrix for the hypergeometric  function ${}_3\mathcal{F}_2(a_0,a_1,a_2,b_1,b_2;y)$ is 
\begin{align}
\label{h3f2}
\bH_{\underline{\mu}}=\left(
\begin{smallmatrix}
 -\frac{\frak{s} \left( 
   b_1\right) \frak{s}  \left(  b_2\right) }{4\, \frak{s}  \left(  a_1\right) \frak{s}  \left(  a_2\right)\frak{s}  \left(a_1-b_1\right)
   \frak{s} \left(a_2-b_2\right) }  & 0 & 0 \\
 0 & -\frac{ \frak{s} \left(b_1\right) \frak{s} \left(
   b_1-b_2\right)}{4\, \frak{s}  \left(  a_0\right) \frak{s}\left(a_0-b_1\right) \frak{s} 
     \left(a_2-b_1\right) \frak{s}  \left(a_2-b_2\right)}  &
   0 \\
 0 & 0 & \frac{\frak{s} 
   \left(b_1-b_2\right) \frak{s} \left(  b_2\right)}{4 \, \frak{s} \left(  a_0\right) \frak{s} 
   \left(a_1-b_1\right)\frak{s}   \left(a_0-b_2\right) \frak{s} \left(a_1-b_2\right)} \\
\end{smallmatrix}
\right)\, .
\end{align} 
 We verified that this intersection matrix $\bH_{\underline{\mu}}$ together with the period matrix $\bP_{\underline{\mu}}$ and the intersection matrix $\bC_{\underline{\mu}}$ fulfils the twisted Riemann bilinear relations in eq.~\eqref{EQU.24} numerically.
This agrees with the intersection matrix that was obtained in ref.~\cite{goto2014intersection} by considering each cycle in an appropriate sub-projective space and then projecting back to the full space. 

\underline{The case $p>2$:} Generalising from eq.~\eqref{h32} we find for the non-zero entries of the homology intersection matrix for $p>2$:
\beq\bsp
\label{genh}
    \left(\bH_{\underline{\mu}}\right)_{11} &=(-2i)^{-p}\prod_{i=1}^p\frac{\frak{s}(b_i) }{\frak{s}(a_i)\frak{s}(a_i-b_i)}\, , \\
    \left(\bH_{\underline{\mu}}\right)_{kk} &= -(-2i)^{-p}\frac{\frak{s}(b_{k-1}) }{\frak{s}(a_0)\frak{s}(a_0-b_{k-1})}\prod_{i=1, i \neq k-1}^p\frac{\frak{s}(b_{k-1}-b_i)}{\frak{s}(b_{k-1}-a_{i})\frak{s}(a_{i}-b_{i})}\,  \, \\
    \left(\bH_{\underline{\mu}}\right)_{ik} &=0 \text{ for } i \neq k
\esp\eeq
This is equivalent to the result of ref.~\cite{goto2014intersection}. We checked numerically that the twisted Riemann bilinear relations are fulfilled for $\bH_{\underline{\mu}}$ together with the periods in eq.~\eqref{perfp} and the cohomology intersection matrix from eq.~\eqref{intF1A} up to $p=7$.


\section{Feynman integrals in one dimension}
\label{app:1D_integrals}

In this appendix we briefly indicate how we can extend Theorem~\ref{twodtheo} to Feynman integrals in $D=1$ dimensions. Using the notations and conventions of section~\ref{introfeyn}, the Feynman integral in one dimension attached to the Feynman graph $G$ reads:
 \begin{align}\label{eq:1Dintegral}
 &\tI_G^1(\underline{\nu},\underline{y}) = \pi^{-L/2}\int_{\mathbb{R}^L} \left(\bigwedge_{j=1}^L  \rd x_j\right)\prod_j\frac{1}{|L_{j}(\underline{x},\underline{y})|^{2\nu_j}} \, ,
  \end{align}
  where the linear forms $L_{j}(\underline{x},\underline{y})$ were defined in eq.~\eqref{eq:props_2D}. The main difference to the integral in eq.~\eqref{EQU.762} lies in the fact that the variables $\underline{x}$ and $\underline{y}$ are real rather than complex.
  
  Note that the absolute values in the propagators in eq.~\eqref{eq:1Dintegral} are essential. If the absolute values
  were absent, then the integrand of eq.~\eqref{eq:1Dintegral} could be identified with the $(L,0)$ form $\omega_G$ from eq.~\eqref{omegag} (with $\nu_j$ replaced by $2\nu_j$). Due to the presence of the absolute values, it is not obvious that the integrand of eq.~\eqref{eq:1Dintegral} fits into the framework of twisted cohomology. In this appendix we argue that this is nonetheless possible.
  
 \begin{quote}
 \begin{thm}
\label{onedtheo}
\emph{Every $L$-loop Feynman graph $G$ with massless propagators in $D=1$ dimensions determines a twisted cohomology group} $H_{\text{dR}}^L(X_G^{\mathbb{R}},\nablasub{-2\underline{\nu}})$\emph{. The value of }$\tI_G^1$ \emph{is then a linear combination of (holomorphic, multi-valued) Aomoto-Gelfand hypergeometric functions.}
\end{thm}
\end{quote}

Theorem~\ref{onedtheo} is the analogue of Theorem~\ref{twodtheo}, but for integrals in one dimension. The twisted cohomology group is almost the same as in two dimensions, the only difference being that the the exponents in the twist are multiplied by 2, and $X_G^{\mathbb{R}} = X_G\cap \mathbb{R}^L$, with $X_G$ defined in eq.~\eqref{eq:X_G_def}. Nevertheless, Theorem~\ref{onedtheo} is somewhat weaker than Theorem~\ref{twodtheo}. Indeed, while in two dimensions we can identify the value of the Feynman integral as the single-valued analogue of a concrete Aomoto-Gelfand hypergeometric function, in one dimension we can only say that it is a linear combination, but we are currently not aware of an a priori way of fixing the coefficients of the linear combination. Note that, just like in two dimensions, one can show that all one-loop integrals in one dimensions are linear combinations of Lauricella $F_D^{(r)}$ functions, while all $L$-loop ladder integrals are linear combinations of generalised hypergeometric ${}_{p+1}F_p$ functions.

Let us now sketch the proof of Theorem~\ref{onedtheo}. From eq.~\eqref{eq:X_G_def} it is easy to see that $X_G^{\mathbb{R}}$ can be written as a disjoint union of (both bounded and unbounded) chambers:
\beq
X_G^{\mathbb{R}} = \bigcup_j \triangle_j\,,\qquad \triangle_i\cap\triangle_j=\emptyset\,,\textrm{~~~if~~~} i\neq j\,.
\eeq
On each chamber, the linear forms $L_{j}(\underline{x},\underline{y})$ are non-zero have a definite sign. As a consequence, we can write
\beq
|L_{j}(\underline{x},\underline{y})|^{2\nu_j}\big|_{\Delta_k} = e^{2i\pi \nu_j\alpha_{jk}}\,L_{j}(\underline{x},\underline{y})\,.
\eeq
The phase factor depends on the Feynman graph under consideration, and we are not aware of an a priori way of determining it. Nevertheless, we can conclude that we can write
\beq
I_G^1(\underline{x},\underline{y}) = \sum_k \left(\prod_je^{2i\pi \nu_j\alpha_{jk}}\right)\,\pi^{-L/2}\,\int_{\triangle_k}\omega_G\,,
\eeq
where $\omega_G$ is the differential form defined in eq.~\eqref{omegag}, but with $\underline{\nu}$ replaced by $2\underline{\nu}$.
The integrals appearing this sum can be evaluated in terms Aomoto-Gelfand hypergeometric functions for $H_{\text{dR}}^L(X_G^{\mathbb{R}},\nablasub{-2\underline{\nu}})$. Indeed, $\omega_G$ defines a cohomology class in $H_{\text{dR}}^L(X_G^{\mathbb{R}},\nablasub{-2\underline{\nu}})$. The corresponding homology group is $H_L(X_G^{\mathbb{R}},\cL_{-2\underline{\nu}})$, and has as a basis the subset of \emph{bounded} chambers. The unbounded chambers can be decomposed into a basis $|\gamma_l]$ of bounded chambers:
\beq
|\Delta_{k}] = |\gamma_l]\,\big(\bH^{-1}_{-2\underline{\nu}}\big)_{ml}\,[\check{\gamma}_m|\triangle_k]\,,
\eeq
where $[\eta_m|$ is a basis of the dual homology group. Hence, we see that each term in the sum is an Aomoto-Gelfand hypergeometric functions:
\beq
I_G^1(\underline{x},\underline{y}) = \sum_k \left(\prod_je^{2i\pi \nu_j\alpha_{jk}}\right)\,\pi^{-L/2}\,\langle\omega_G|\gamma_l]\big(\bH^{-1}_{-2\underline{\nu}}\big)_{ml}\,[\check{\gamma}_m|\triangle_k]\,.
\eeq

\bibliographystyle{jhep}
\bibliography{twistedbib}

\end{document}